\documentclass[oneside,12pt]{article}

\usepackage{color}
\usepackage{float}
\usepackage{amsmath,amssymb}
\usepackage{mathrsfs}
\usepackage{enumerate}
\usepackage{breqn}
\usepackage{bm}
\usepackage{harpoon}
\usepackage{mathtools}
   \mathtoolsset{showonlyrefs}
\usepackage{graphicx}
\usepackage{epsfig,psfrag,subfig}
\usepackage{hyperref}
\hypersetup{bookmarks=false, colorlinks=true, linkcolor=black, urlcolor=black, pagebackref=true, pdfstartview={FitH}, bookmarksopenlevel=2 }

\usepackage{epstopdf}
\usepackage{import}
\graphicspath{{Figures/}}

\DeclareMathAlphabet{\mathbf}{OT1}{cmr}{bx}{it}

\definecolor{red}{rgb}{0.9,0,0}
\definecolor{blue}{rgb}{0.019,0.506,0.62}
\definecolor{green}{rgb}{0.0,0.5,0.2}
\definecolor{darkblue}{rgb}{0.2,0.2,0.5}
\definecolor{orange}{rgb}{0.96,0.45,0.184}

\newcommand{\tO}[1]{{\color{orange} #1}}

\newcommand{\bydef}{\,\raise.050ex\hbox{\rm:}\kern-.025em\hbox{\rm=}\,}
\newcommand{\defby}{=\raise.075ex\hbox{\kern-.325em\hbox{\rm:}}\,}

\def\qed{\relax\ifmmode\hskip2em \Box\else\unskip\nobreak\hskip1em $\Box$\fi}
\newcommand {\Cc}  {\mathcal{C}}

\newcommand {\Pc}  {\mathcal{P}}

\newcommand {\Tc}  {\mathcal{T}}
\newcommand {\Uc}  {\mathcal{U}}

\newcommand {\Zc}  {\mathcal{Z}}
\newcommand {\ab} {\mathbf{a}}

\newcommand {\db} {\mathbf{d}}
\newcommand {\eb} {\mathbf{e}}

\newcommand {\gb} {\mathbf{g}}
\newcommand {\hb} {\mathbf{h}}

\newcommand {\pb} {\mathbf{p}}

\newcommand {\xb} {\mathbf{x}}
\newcommand {\yb} {\mathbf{y}}

\newcommand {\zb} {\mathbf{z}}

\newcommand {\Ro} {\mathbb{R}}

\DeclareMathOperator{\Thz}{\overset{(z)}{\Theta}}
\DeclareMathOperator{\Thc}{\overset{(c)}{\Theta}}
\DeclareMathOperator{\Ths}{\overset{(s)}{\vartheta}}

\DeclareMathOperator{\pcl2}{\overset{c}{p}_{2\ell}}
\DeclareMathOperator{\pc2}{\overset{c}{p}_{2}}
\DeclareMathOperator{\psl}{\overset{s}{p}_{\ell}}
\DeclareMathOperator{\ps}{\overset{s}{p}}
\DeclareMathOperator{\qcl}{\overset{c}{q}_{\ell}}
\DeclareMathOperator{\qsl}{\overset{s}{q}_{\ell}}

\newcommand\jump[1]{[\![#1]\!]}

\newtheorem{thm}{Theorem}

\newtheorem{lem}[thm]{Lemma}

\newtheorem{proof}[thm]{Proof}

\newtheorem{rem}{Remark}

\def\proof{{\noindent{\sc Proof.\ }}}
\newcommand{\QED}{\hspace{1ex}\hfill$\Box$\vspace{2ex}}

\begin{document}

\title{\vspace{-1cm} \textbf{A REBO-potential-based 
		model for\\ graphene bending  by $\boldsymbol{\Gamma}$-convergence}}

\author{
Cesare Davini$^1$  \!\!\!\!\! \and Antonino Favata$^2$  \!\!\!\!\! \and Roberto Paroni$^3$ 
}

\date{\today}

\maketitle

\vspace{-1cm}
\begin{center}
	{\small
		$^1$ Via Parenzo 17, 33100 Udine\\
		\href{mailto:cesare.davini@uniud.it}{cesare.davini@uniud.it}\\[8pt]
		$^2$ Department of Structural and Geotechnical Engineering\\
		Sapienza University of Rome, Rome, Italy\\
		\href{mailto:antonino.favata@uniroma1.it}{antonino.favata@uniroma1.it}\\[8pt]

		$^3$ DADU\\
		University of Sassari, Alghero (SS), Italy\\
		\href{mailto:paroni@uniss.it}{paroni@uniss.it}
	}
\end{center}

\pagestyle{myheadings}
\markboth{C.~Davini, A.~Favata, R.~Paroni }
{A REBO-potential-based 
	model for  graphene bending  by $\Gamma$-convergence}

\vspace{-0.5cm}
\section*{Abstract}
An atomistic to continuum model for a graphene sheet undergoing bending is presented. Under the assumption that the atomic interactions are governed by a harmonic approximation of the 2nd-generation Brenner REBO  (reactive empirical bond-order) potential, involving  first, second and third nearest neighbors of any given atom, we determine the variational limit  of the energy functionals. It turns out that the $\Gamma$-limit depends on the linearized mean and Gaussian curvatures.  If some specific contributions in the atomic interaction are neglected, the variational limit  is non-local.

\vspace{0.5cm}
\noindent \textbf{Keywords}: Graphene bending, Homogenization, $\Gamma$-convergence, Non-locality

\tableofcontents

\section{Introduction}\label{sec:INTROD}

Graphene is a two-dimensional carbon allotrope, in the form of a hexagonal lattice whose vertices are occupied by C atoms. It has recently attracted a huge interest of the scientific community, due to   its extraordinary mechanical, electrical and thermal conductivity properties \cite{Akiwande_2016}, that make graphene a candidate for a great variety of technological applications; actually, its potentialities, and those of graphene-based materials, are far from being fully understood, and many studies are carried out  in order to develop new technological applications \cite{Ferrari2014}. In particular, understanding the bending  behavior of graphene represents a challenge of significant interest because of possible  applications in the field of flexible devices.

For the modeling of graphene many different approaches at different scales can be found in the literature, ranging from first principle calculations \cite{Kudin_2001,Liu_2007}, atomistic calculations \cite{Zakharchenko_2009,Zhao_2009,SakhaeePour_2009} and continuum mechanics \cite{Yakobson_1996,Cadelano_2009,Luhuang2009,Scarpa_2009,Scarpa_2010,Sfyris_2014b,Sfyris_2014}; furthermore, mixed atomistic formulations with finite elements have been reported for graphene \cite{Arroyo_2002,Arroyo2004}.  Both in-plane and bending deformations  have been studied in \cite{Luhuang2009} and the out-of-plane bending behavior has been investigated in \cite{Scarpa_2009,Scarpa_2010} with the use a special equivalent atomistic-continuum model. In \cite{Zhou_2008}, the elastic properties of graphene have been theoretically predicted on taking into account internal lattice relaxation. In \cite{Cadelano_2009}, by combining continuum elasticity theory and tight-binding atomistic simulations, a constitutive nonlinear stress-strain relation for graphene stretching has been proposed. Atomistic simulations have been employed to investigate the elastic properties of graphene in \cite{SakhaeePour_2009}. Based on the experiments performed in \cite{Lee2008}, the nonlinear in-plane elastic properties of graphene have been calculated in \cite{Wei_2013} by means of DFT. A continuum theory of a free-standing graphene monolayer, viewed as a two dimensional 2-lattice, has been proposed in \cite{Sfyris_2014b,Sfyris_2014}, where the shift vector which connects the two simple lattices is considered as an auxiliary variable. 

When a continuum picture is pursued, the key point of modeling relies in the connection between the atomistic  and the gross description. Frequently, the target continuum model is postulated and that connection is established through  a suitable choice of constitutive and geometric parameters. 

In this paper, the connection is set within the general framework of homogenization theory.  For the case of the out-of-plane deformations of graphene, we determine the variational limit ---in the sense of $\Gamma$-convergence---  of the discrete energy functionals under a topology that guarantees the convergence of minimizers. Thus,  the limit functional describes a continuous two-dimensional medium fully accounting for the bending behavior of a graphene sheet. 

Homogenization of graphene has already been studied in \cite{LeDret_2011,LeDret_2013,Davini_2014}. In these works the membranal equations have been deduced, non-linear in \cite{LeDret_2011,LeDret_2013}  and linearized in \cite{Davini_2014}; moreover, interactions up to the second neighbor have been taken into account.

Our description of  the atomic-scale interaction is based on the discrete mechanical model proposed in \cite{Favata_2016} and exploited in  \cite{Favata_2016a,Favata_2016b,Alessi_2016,Davini_2017,Davini_2017a}, where the results are also obtained for the  2nd-generation Brenner REBO  (reactive empirical bond-order)  potential, which is largely used in Molecular Dynamics simulations for carbon allotropes. Here, we recall the most relevant features:
\begin{enumerate}[(i)]
	\item interatomic bonds  involve first, second and third nearest neighbors of any given atom. In particular, the kinematical variables we consider are \textit{bond lengths}, \textit{bond angles}, and \textit{dihedral angles}; from \cite{Brenner_2002} it results that these latter are of two kinds, that we here term C and Z, as  described in Sec. \ref{sec:KIN}.
	\item graphene does not have a configuration at ease. In particular an angular \textit{self-stress} is present,  and the \textit{self-energy} associated with the self-stress (sometimes called \textit{cohesive energy} in the literature)  needs to be considered.
\end{enumerate}
Resting on this atomistic energetic description, we here determine the equivalent continuum   limit. A pointwise limit has already been determined in \cite{Davini_2017}; we here prove a compactness result and determine the $\Gamma$-limit, which in turn guarantee the convergence of minima and minimizers.  The main results we obtain are:
\begin{enumerate}[(i)]
	\item The $\Gamma$-limit  energy  depends on the square of the mean curvature and on the Gaussian curvature; the constitutive coefficients depend on the dihedral contribution and  the self-stress.
	\item If both the self-stress and the C-energy are neglected, the $\Gamma$-limit is non-local and depends on a function which is solution of a differential problem. 	
\end{enumerate}

That graphene could be modeled in the framework of the non-local elasticity has been conjectured in the literature several times. A  review of recent research studies on this matter can be found in \cite{Liew_2017}. Unlike classical continuum models, within  the framework of non-local elasticity  it is  assumed that the stress at a reference point in a body depends not only on the strains at that point, but also on strains at all other points of the body. Since classical model are inefficient to model the mechanical  behavior of graphene, and since the description of the atomic bonds leads to consider relatively long interactions, some authors have believed reasonable to \textit{postulate} a kind of non-locality in the model. As a matter of fact, the connection between atomistic and continuum description has never been mathematically rigorous, and has been limited to fit additional parameters to experiments or atomistic simulations.    

The paper is organized as follows. In Sec. \ref{sec:KIN} we present a description of graphene energetics at atomistic level, as suggested by the 2nd generation Brenner potential.  In Sec.  \ref{results} we lay down our  assumptions and announce the main results. In Sec. \ref{dihe} we introduce some interpolating functions and determine their limits when the lattice size tends to zero. In Sec. \ref{sec:Zdihedr} we determine lower bounds of the limit energy and prove a theorem concerning the regularity of the limit function. In Sec. \ref{proof23} we determine the $\Gamma$-limit for the general case and the case when self-stress and C-energy are neglected.

\vspace{0.2cm}


\noindent
{\bf Notation.}
We use the direct notation. We denote vectors by  low-case Roman bold-face  letters and scalar fields by  low-case Roman or Greek light-face letters. The canonical basis for $\Ro{^3}$  is denoted by $\{\eb_1,\eb_2,\eb_3\}$. For a vector $\ab$ we set $\ab^\perp=\eb_3\wedge \ab$, the vector $\ab$ rotated by $\pi/2$ counter-clockwise.  For a given scalar field $w$, we denote by $\nabla w$ its gradient and by 
$$\partial_\ab w:=\nabla w\cdot \frac{\ab}{|\ab|}$$
the  derivative in direction $\ab/|\ab|$.

\section{The bending energy of a graphene sheet} \label{sec:KIN}

\begin{figure}[h!]
	\begin{center}
		\def\svgwidth{.7\textwidth}
		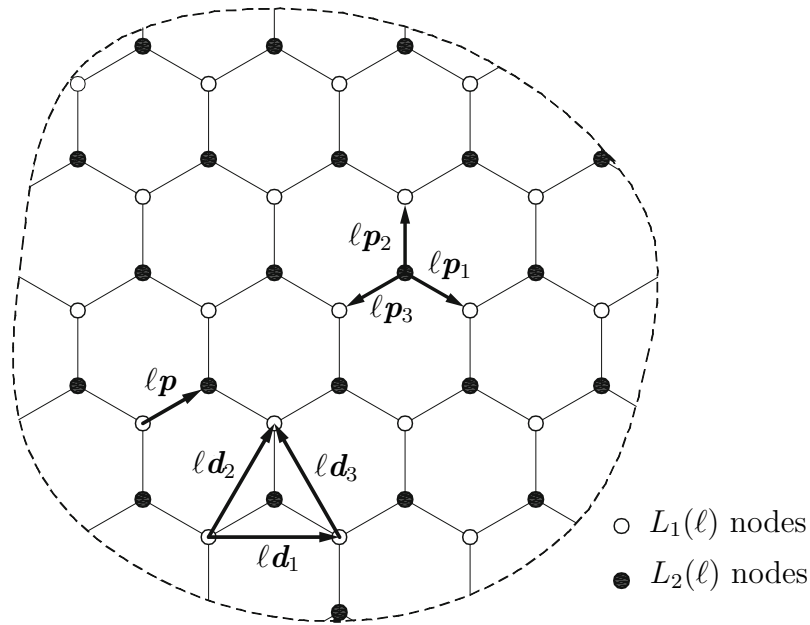
		\caption{The hexagonal lattice}\label{fg:LATTICE}
	\end{center}
\end{figure}

As reference configuration we use the $2$--{\it lattice} generated by two simple Bravais lattices 
\begin{equation}\label{eq:KIN_1}
\begin{array}{l} L_1(\ell) = \{ \mathbf{x} \in \mathbb{R}^2:
\mathbf{x} = n^1\ell\db_{1} + n^2\ell \db_{2} \quad \mbox{with} \quad (n^1,
n^2) \in \mathbb{Z}^2 \}, \\ L_2(\ell) = \ell\mathbf{p} + L_1(\ell),
\end{array}
\end{equation}
simply shifted with respect to one another by $\ell\pb$, see
Fig.~\ref{fg:LATTICE}.
In \eqref{eq:KIN_1}, $\ell$ denotes
the lattice size (the reference {\it interatomic distance}), while
$\ell\db_{\alpha}$ and $\ell\mathbf{p}$  respectively are the {\em lattice
	vectors} and the {\em shift vector}, with

\begin{equation}\label{eq:KIN_2}
\db_{1} =\sqrt{3}\eb_1, \quad \db_{2} = \frac{\sqrt{3}}{2}\eb_1+
\frac{3}{2}\eb_2 \quad \mbox{and} \quad \mathbf{p} =
\frac{\sqrt{3}}{2}\eb_1+ \frac{1}{2}\eb_2.
\end{equation}
The sides of the hexagonal cells in Figure~\ref{fg:LATTICE} stand for the bonds
between pairs of next nearest neighbor atoms and are represented by the
vectors
\begin{equation}\label{eq: KINENER 3}
\pb_\alpha = \db_\alpha - \pb \ \  (\alpha = 1, 2) \quad \mbox{and}
\quad \pb_3 = - \, \pb.
\end{equation}
For convenience we also set $$\db_3=\db_2-\db_1.$$

In what follows we denote by
\begin{equation}\label{eqnew}
\xb^\ell=n^1\ell\db_1+n^2\ell\db_2+m\ell\pb, \quad(n^1,n^2,m)\in\mathbb{Z}^2\times\{0,1\}
\end{equation}
the lattice points and label them by the triplets $(n^1,n^2,m)$: the points with $m=0$ belong to $L_1(\ell)$, while those in $L_2(\ell)$ correspond to $m=1$.

\begin{figure}
	\begin{center}
		\def\svgwidth{.9\textwidth}
		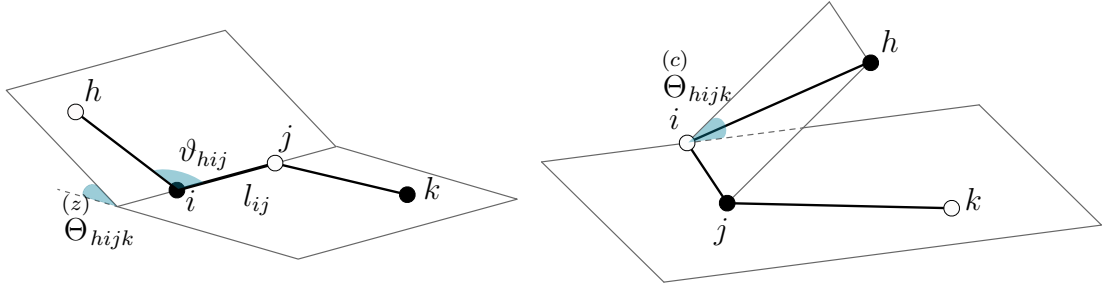
		\caption{Edge bond $l$, wedge bond $\vartheta$,  Z-dihedron  $\Thz$ and a C-dihedron $\Thc$.}
		\label{fig:bonds}
	\end{center}
\end{figure}

Graphene energetics depends on the description chosen to mimic atomic  interactions. Our model stems on the 2nd-generation Brenner potential \cite{Brenner_2002}, which is one of the most used in molecular dynamics simulations of graphene. Accordingly, the  binding energy $V$ of an atomic aggregate is given as a sum over nearest neighbors:
\begin{equation}\label{V}
V=\sum_i\sum_{j<i} V_{ij}\,, \quad V_{ij}=V_R(l_{ij})+b_{ij}(\vartheta_{hij},\Theta_{hijk})V_A(l_{ij}),
\end{equation}
where the individual effects of the \emph{repulsion} and \emph{attraction functions} $V_R(l_{ij})$ and $V_A(l_{ij})$, which model pair-wise interactions of  atoms $i$ and $j$ depending on their distance $l_{ij}$, are modulated by the \emph{bond-order function} $b_{ij}$; for a given bond chain of atoms $h,i,j,k$ the function $b_{ij}$ depends in a complex manner on the angle between the edges $hi$ and $ij$ and on the dihedral angle between the planes spanned by $(hi,ij)$ and $(ij,jk)$. This potential reveals that,  in order to properly account  for the mechanical behavior of  graphene, it is necessary to consider three types of energetic contributions: binary interactions between next nearest atoms  ({\it edge bonds}), three-bodies interactions between consecutive pairs of next nearest atoms  ({\it wedge bonds}) and four-bodies interactions between three consecutive pairs of next nearest atoms ({\it dihedral bonds}).  There are two types of relevant dihedral bonds: the Z-dihedra, in which the edges connecting the four atoms form a Z-shape, and the C-dihedra, in which the edges form a C-shape (see Fig. \ref{fig:bonds}).

Following \cite{Davini_2017}, we  consider a harmonic approximation of the  energy density.
Moreover, it is possible to show \cite{Favata_2016} that the edge length at ease is $\ell$, the dihedral angle at ease is null, while  the angle  at ease between consecutive edges is $\frac{2}{3}\pi + \delta\vartheta_0$, where $\delta\vartheta_0\ne 0$. This means that the graphene sheet does not have a configuration at ease ({\it i.e.} stress-free).

With this in mind, we assume that the energy  is given by the sum of the following terms:

\begin{equation}\label{eq:ENER 1bis}
\begin{array}{l}
\displaystyle \mathcal{U}_\ell^l = \frac{1}{2} \, \sum_{\mathcal {E}} k^l \, (\delta l)^2 ,\\
\displaystyle \mathcal{U}_\ell^\vartheta = \tau_{0} \sum_{\mathcal {W}}\delta \vartheta +  \frac{1}{2} \, \sum_{\mathcal {W}} k^\vartheta \, (\delta\vartheta)^2,\\
\displaystyle\mathcal{U}_\ell^\Theta = \frac{1}{2} \, \sum_{\mathcal {Z}} k^{\Zc}\, (\delta\Thz)^2+\frac{1}{2} \, \sum_{\mathcal {C}} k^{\Cc}\, (\delta\Thc)^2.
\end{array}
\end{equation}
$\mathcal{U}_\ell^l$, $\mathcal{U}_\ell^\vartheta $ and $\mathcal{U}_\ell^\Theta$ are the energies of the edge bonds, the wedge bonds and the dihedral bonds, respectively; $\delta l$ denotes the change of distance between nearest neighbor atoms, $\delta\vartheta$ the change of angle between pairs of edges having a lattice point in common and $\delta\Thz$ and $\delta\Thc$ the Z- and C-dihedral change of angles between two consecutive wedges; finally,
\begin{equation}\label{deft0}
\tau_0 := -k^\vartheta \, \delta\vartheta_0
\end{equation} 
is the {\it wedge self-stress}. The sums extend to 	all edges, $\mathcal{E}$, all wedges, $\mathcal{W}$, all Z-dihedra, $\mathcal{Z}$, and all C-dihedra $\mathcal{C}$. 	The bond constants $k^l$, $k^\vartheta$, $k^\Zc $,  and $k^\Cc$  can be deduced by making use of the 2nd-generation Brenner potential.
In \eqref{eq:ENER 1bis}  we  approximate the strain measures to the lowest order that makes the energy quadratic in the displacement field. 

In \cite{Davini_2017} we have shown that the  change in length of edges and the first order  variation of the  change in angle of wedges depend on the in-plane components of the displacement. 
We have also shown that the energy splits into two contributions: one depends on the in-plane displacement and the other ---the bending energy--- is a function of  the out-of-plane displacement $w: L_1(\ell) \cup L_2(\ell)\to\mathbb{R}$.
The total bending energy associated to  $w$ is  given by
\begin{equation}\label{bend 2}
\mathcal{U}_\ell (w)=  \mathcal{U}^{\mathcal{Z}}_\ell(w)+ \mathcal{U}^{\mathcal{C}}_\ell(w)+\mathcal{U}^s_\ell(w),
\end{equation}
where 
\begin{equation}\label{encontr}
\begin{aligned}
& \mathcal{U}^{\mathcal{Z}}_\ell(w):=\frac{1}{2} \, \sum_{\mathcal {Z}} k^\Zc \, (\delta\Thz)^2\\ & \mathcal{U}^{\mathcal{C}}_\ell(w):=\frac{1}{2} \, \sum_{\mathcal {C}} k^\Cc \, (\delta\Thc)^2,\\
&\Uc_\ell^{s} :=\tau_{0} \sum_{\mathcal {W}}\delta \vartheta^{(2)}.\\ 
\end{aligned}
\end{equation}
$\Uc_\ell^{\Zc}$ and $\Uc_\ell^{\Cc}$ are the \textit{Z-} and \textit{C-dihedral energy}, while $\Uc_\ell^{s}$ is the \textit{self-energy} (corresponding to the so-called \textit{cohesive energy} in the literature); in $\eqref{encontr}_3$, $\delta\vartheta^{(2)}$ is the  second order variation of the wedge angle with respect to the reference angle $\frac{2}{3}\pi$. 

\begin{figure}
	\begin{center}
		\def\svgwidth{.8\textwidth}
		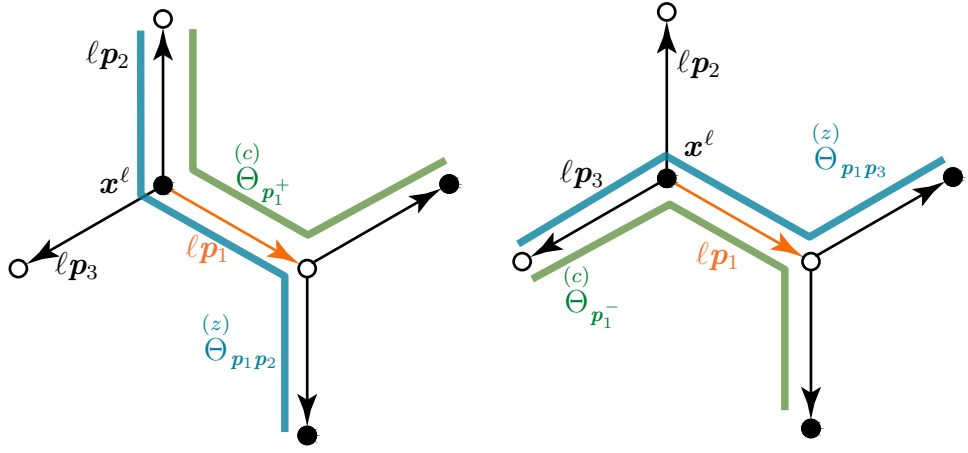
		\caption{Left: C-dihedral angles $\Thc_{\pb_1^+}$ (green) and Z-dihedral angle $\Thz_{\pb_1\pb_2}$ (blue). Right: C-dihedral angles $\Thc_{\pb_1^-}$ (green) and Z-dihedral angle $\Thz_{\pb_1\pb_3}$ (blue).}
		\label{fig:cell_text}
	\end{center}
\end{figure}

Hereafter we write explicitly the dependence  on $w$ of the strain measures. In particular, by $\Thz_{\pb_i\pb_{i+1}}[w](\xb^\ell)$ we denote the Z-dihedral angle, associated to displacement $w$, that corresponds 
to the Z-dihedron with middle edge $\ell \pb_i$, starting from $\xb^\ell$ and the other two edges parallel to $\pb_{i+1}$. The  C-dihedral angle $\Thc_{\pb_i^+}[w](\xb^\ell)$ is the angle corresponding to the C-dihedron with middle edge $\ell \pb_i$ and oriented as $\pb_i^\perp$,  while $\Thc_{\pb_i^-}[w](\xb^\ell)$ is the angle corresponding to the C-dihedron oriented opposite to $\pb_i^\perp$ (see Fig.~\ref{fig:cell_text} for $i=1$). 

In \cite{Davini_2017} we have shown that the Z-dihedral energy has the following expression:
\begin{equation}\label{enZ}
\mathcal{U}^{\mathcal{Z}}_\ell(w):=\frac{1}{2}k^{\mathcal{Z}} \, \sum_{\xb^\ell\in L_2(\ell)} \sum_{i=1}^3
\Big(\Thz_{\pb_i\pb_{i+1}}[w](\xb^\ell)\Big)^2+ \Big(\Thz_{\pb_i\pb_{i+2}}[w](\xb^\ell)\Big)^2,
\end{equation}
where the change of Z-dihedral angles may be given in the following way:
\begin{equation}\label{Z-wedge}
\begin{aligned}
\Thz_{\pb_i\pb_{i+1}}[w](\xb^\ell)&=
\frac{2\sqrt{3}}{3\ell}[w(\xb^\ell+\ell\pb_i-\ell \pb_{i+1})-w(\xb^\ell+\ell \pb_{i})+w(\xb^\ell+\ell \pb_{i+1})\\
&\hspace{7.5cm}-w(\xb^\ell)],\\
\Thz_{\pb_i\pb_{i+2}}[w](\xb^\ell)&=\frac{2\sqrt{3}}{3\ell}[w(\xb^\ell+\ell\pb_i-\ell \pb_{i+2})-w(\xb^\ell+\ell \pb_{i})+w(\xb^\ell+\ell \pb_{i+2})\\
&\hspace{7.5cm}-w(\xb^\ell)].
\end{aligned}
\end{equation}
To make notation simpler,
we have omitted the symbol $\delta$ to denote the variation; we will do the same  throughout the paper without any further mention.
Analogously, the C-dihedral energy can be written as:
\begin{equation}\label{enC}
\begin{aligned}
\mathcal{U}^{\mathcal{C}}_\ell(w):=\frac{1}{2} k^{\mathcal{C}} \, \sum_{\xb^\ell\in L_2(\ell)}
\sum_{i=1}^3 \Big(\Thc_{\pb_i^+}[w](\xb^\ell)\Big)^2+ \Big(\Thc_{\pb_i^-}[w](\xb^\ell)\Big)^2,
\end{aligned}
\end{equation}
with the change of C-dihedral angles given by:
\begin{equation}\label{C-wedge}
\begin{aligned}
\Thc_{\pb_i^+}[w](\xb^\ell)&=+\frac{2\sqrt{3}}{3\ell}[2w(\xb^\ell)-w(\xb^\ell+\ell \pb_{i+1})+w(\xb^\ell+\ell\pb_i-\ell \pb_{i+2})\\
&\hspace{7cm}-2w(\xb^\ell+\ell \pb_{i})],\\
\Thc_{\pb_i^-}[w](\xb^\ell)&=-\frac{2\sqrt{3}}{3\ell}[2w(\xb^\ell)-w(\xb^\ell+\ell \pb_{i+2})+w(\xb^\ell+\ell\pb_i-\ell \pb_{i+1})\\
&\hspace{7cm}-2w(\xb^\ell+\ell\pb_i)].
\end{aligned}
\end{equation}
Concerning the self-energy,  \eqref{encontr}$_3$ becomes:
\begin{equation}\label{ens}
\mathcal{U}^{s}_\ell(w):=-\frac{1}{2} \tau_0\,  \Big[\sum_{\xb^\ell\in L_1(\ell)} 
\, \Big(\Ths_1[w](\xb^\ell)\Big)^2+\sum_{\xb^\ell\in L_2(\ell)} 
\,\Big( \Ths_2[w](\xb^\ell)\Big)^2\Big],
\end{equation}
where:
\begin{equation}\label{s_angles}
\begin{aligned}
\Ths_1[w](\xb^\ell)&=\frac{\sqrt{3\sqrt{3}}}{\ell}\Big(\frac 13 \sum_{i=1}^3w(\xb^\ell-\ell\pb_i)-w(\xb^\ell)\Big), \quad \xb^\ell\in L_1(\ell),\\
\Ths_2[w](\xb^\ell)&=\frac{\sqrt{3\sqrt{3}}}{\ell}\Big(\frac 13 \sum_{i=1}^3w(\xb^\ell+\ell\pb_i)-w(\xb^\ell)\Big), \quad \xb^\ell\in L_2(\ell).
\end{aligned}
\end{equation}
The reader is referred to \cite{Davini_2017} for detailed computations.

\section{Main assumptions and results}\label{results}

\begin{figure}[h]
	\centering
	\def\svgwidth{.6\textwidth}
	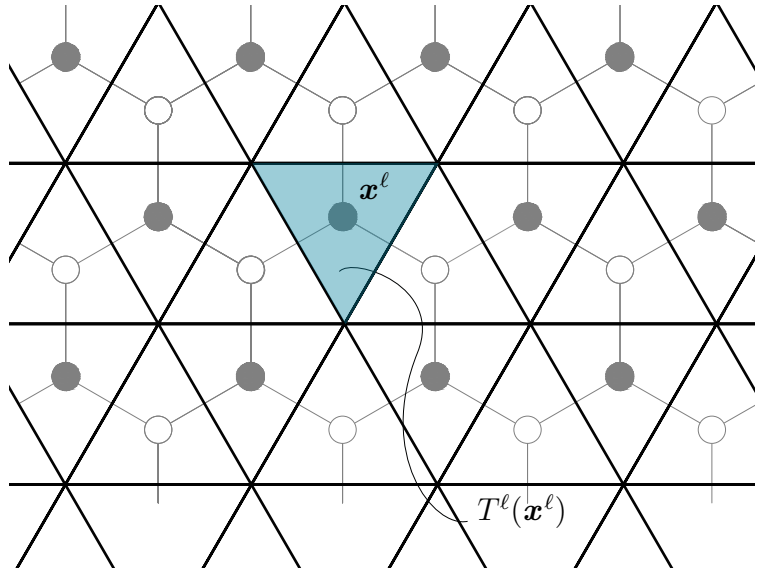
	\caption{Dual triangulation.}\label{triangulation}
\end{figure}

The dual triangulation represented in Fig.~\ref{triangulation}
will play an important role in our analysis. This is composed by equilateral triangles of length side $\sqrt{3}\ell$ 
centered at the lattice points that tessellate $\mathbb{R}^2$. The triangle centered at point $\xb^\ell$ is denoted by $T^\ell(\xb^\ell)$. 

In the place of displacements defined  on the lattice points of $L_1(\ell) \cup L_2(\ell)$,
as introduced in the previous section,
we shall consider (equivalent) functions with domain $\Ro^2$ that are constant
over each triangle of the dual triangulation. 
Thence, if $w$ is such a function the following representation holds:
\begin{equation}\label{defw}
w(\xb)=\sum_{\xb^\ell\in L_1(\ell)\cup L_2(\ell)}w(\xb^\ell)\ \chi_{T^\ell(\xb^\ell)}(\xb),
\end{equation}
where $\chi_{T^\ell(\xb^\ell)}$  is the characteristic function of $T^\ell(\xb^\ell)$.
The energies and the deformation measures defined in the previous section can
be unambiguously evaluated on this kind of functions.

We consider a graphene sheet of finite extension
corresponding to the nodes of lattice $L_1(\ell) \cup L_2(\ell)$ contained in  
some open and simply connected bounded set $\Omega$ of
$\mathbb{R}^2$. The lattice size $\ell$ is assumed to be much
smaller than the diameter of the largest ball contained in
$\Omega$.
For simplicity, we consider homogeneous Dirichlet boundary conditions, 
which we implement by considering out-of-plane displacements belonging to the set
\begin{align*}
\mathcal{A}_\ell=\{w\in L^2(\Ro^2): \ & w \mbox{ is constant over each $T^\ell(\xb^\ell)$ and}\\
& w=0 \mbox{ over all $T^\ell(\xb^\ell)\nsubseteq \Omega$}\}.
\end{align*}
Before stating our first result we make the following assumption:
\begin{equation}\label{assumption}
k^{\mathcal{Z}}>0,\quad k^{\mathcal{C}}\ge0 \quad\mbox{and } \quad\tau_0\le 0
\end{equation}
that will be maintained throughout the paper without any further mention.

The following compactness result holds.

\begin{thm}\label{thm-wstrong_0}
	Let $w_\ell\in\mathcal{A}_\ell$ be a sequence that satisfy the energy bound:
	\begin{equation}\label{energybound}
	\sup_\ell \mathcal{U}_\ell(w_\ell)<+\infty. 
	\end{equation} 
	Then, there exist $w\in H^2_0(\Omega)$ and a subsequence of $\{w_\ell\}$, 
	not relabelled, such that
	\begin{equation}\label{wlw}
	w_\ell\to w \quad \mbox{in }L^2(\Omega).
	\end{equation}
\end{thm}

All the theorems stated in this section will be proved in the following sections.

The next Theorem characterizes the bending behavior of graphene. 

\begin{thm}\label{Gamma}
	Assume that either $k^{\mathcal{C}}\ne 0$ or $\tau_0\ne 0$, and set
	\begin{equation}\label{Uel}
	\mathcal{U}^{\rm e}_\ell (w):=
	\left\{
	\begin{array}{ll}
	\mathcal{U}_\ell (w) & \mbox{if }w\in \mathcal{A}_\ell,\\
	+\infty & \mbox{if } w\in L^2(\Omega)\setminus \mathcal{A}_\ell.
	\end{array}
	\right.
	\end{equation}
	The functionals $\mathcal{U}^{\rm e}_\ell$ $\Gamma$-converge with respect to the $L^2(\Omega)$-convergence to the functional
	$$
	\mathcal{U}^{\rm e}_0(w):=
	\left\{
	\begin{array}{ll}
	\mathcal{U}^{(b)}_0(w) & \mbox{if }w\in H^2_0(\Omega),\\
	+\infty & \mbox{if } w\in L^2(\Omega)\setminus H^2_0(\Omega),
	\end{array}
	\right.
	$$
	where
	\begin{equation}\label{U0b}
	\begin{aligned}
	\mathcal{U}^{(b)}_0(w)
	:=&\frac 12 \int_\Omega\Bigg(\frac{5\sqrt{3}}3 k^{\mathcal{Z}}+\frac{2\sqrt{3}}{3} k^{\mathcal{C}}-\frac{\tau_0}2\Bigg)( \Delta w)^2\\
	&\hspace{3cm}- \frac{8\sqrt{3}}3 \big(k^{\mathcal{Z}}+ k^{\mathcal{C}}\big)\det \nabla^2 w\,d\xb.
	\end{aligned}
	\end{equation}
\end{thm}

We close this section by looking at the case $k^{\mathcal{C}}=\tau_0= 0.$
For $w\in H^2_0(\Omega)$ we denote by
$$
(-\Delta)^{-1}(-\frac 23 \partial_{\pb_1\pb_2\pb_3}w)\in H^1_0(\Omega)
$$
the solution of the following problem
\begin{equation}\label{invW}
\left\{
\begin{aligned}
&\gamma \in H^1_0(\Omega),\\
&-\Delta \gamma= -\frac 23 \partial_{\pb_1\pb_2\pb_3}w\quad \mbox{in }\mathcal{D}'(\Omega).
\end{aligned}
\right.
\end{equation}
The following theorem  characterizes the $\Gamma$-limit when $k^\Cc=\tau_0=0$, that is  $\Uc_\ell(w)=\mathcal{U}^{\mathcal{Z}}_\ell (w)$. We notice that in this case the $\Gamma$-limit is  a non-local functional.

\begin{thm}\label{Gamma2} Let
	\begin{equation}\label{UZel}
	\mathcal{U}^{\mathcal{Z}\rm e}_\ell (w):=
	\left\{
	\begin{array}{ll}
	\mathcal{U}^{\mathcal{Z}}_\ell (w) & \mbox{if }w\in \mathcal{A}_\ell,\\
	+\infty & \mbox{if } w\in L^2(\Omega)\setminus \mathcal{A}_\ell.
	\end{array}
	\right.
	\end{equation}
	The functionals $\mathcal{U}^{\mathcal{Z}\rm e}_\ell$ $\Gamma$-converge with respect to the $L^2(\Omega)$-convergence to the functional
	$$
	\mathcal{U}^{\mathcal{Z}\rm e}_0(w):=
	\left\{
	\begin{array}{ll}
	\mathcal{U}^{\mathcal{Z}(b)}_0\left(w,(-\Delta)^{-1}(-\frac 23 \partial_{\pb_1\pb_2\pb_3}w)\right) & \mbox{if }w\in H^2_0(\Omega),\\
	+\infty & \mbox{if } w\in L^2(\Omega)\setminus H^2_0(\Omega),
	\end{array}
	\right.
	$$
	where
	\begin{align}
	\mathcal{U}^{\mathcal{Z}(b)}_0(w, \gamma):=&
	\frac{5\sqrt{3}}{6}\,k^{\mathcal{Z}}\int_{\Omega}( \Delta w)^2- \frac{8}5\det \nabla^2 w\,d\xb\nonumber\\
	&+\sqrt{3}\,k^{\mathcal{Z}}\int_{\Omega}\left(2\partial_{12}w\eb_1+(\partial_{11}w-\partial_{22}w)\eb_2\right)\cdot\nabla \gamma\,d\xb\label{Uz00}\\
	&+3\sqrt{3}\,k^{\mathcal{Z}}\int_{\Omega}|\nabla \gamma|^2\,d\xb.\nonumber
	\end{align}
\end{thm}

\section{Interpolating functions and their limits}\label{dihe}

In this section we introduce three piecewise affine interpolants on strips of $\Ro^2$ that are naturally emerging in the study of  the behavior of the $\mathcal{Z}$-dihedral energy. These interpolants, besides shading light on the $\mathcal{Z}$-dihedral energy, will play a crucial role also in the study
of the other energies.

\begin{figure}[h]
	\centering
	\def\svgwidth{.7\textwidth}
	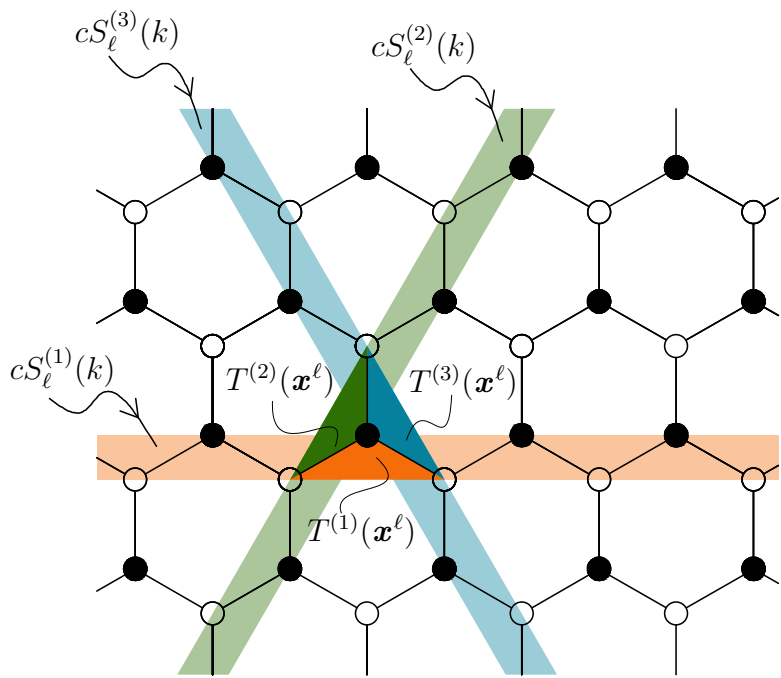
	\caption{A representation of the strips $cS_\ell^{(i)}(k)$ and of the triangles $T^{(i)}(\xb^\ell)$. }\label{strips}
\end{figure}

It is convenient to group the nodes of the $2$--$lattice$ as
\begin{align*}
S_\ell^{(1)}(k)&=\{\xb^\ell\in\mathbb{R}^2: \ \xb^\ell=(i\db_1+k\db_2+m\pb)\ell \  \mbox{with } i\in\mathbb{Z},\  m\in\{0,1\}\}, \\
S_\ell^{(2)}(k)&=\{\xb^\ell\in\mathbb{R}^2: \ \xb^\ell=(k \db_1+i \db_2+m\pb)\ell \  \mbox{with } i\in\mathbb{Z},\  m\in\{0,1\}\} ,\\
S_\ell^{(3)}(k)&=\{\xb^\ell\in\mathbb{R}^2: \ \xb^\ell=((k-i-m) \db_1+i \db_2+m\pb)\ell \  \mbox{with } i\in\mathbb{Z},\\
&\hspace{8cm}  m\in\{0,1\}\}.
\end{align*}
We denote by $cS_\ell^{(i)}(k)$ the convex hull of $S_\ell^{(i)}(k)$, see Fig. \ref{strips}, and define 
$$
S_\ell^{(i)}=\bigcup_{k\in \mathbb{Z}} S_\ell^{(i)}(k), \qquad  cS_\ell^{(i)}=\bigcup_{k\in \mathbb{Z}} cS_\ell^{(i)}(k),\qquad \mbox{for } i=1,2,3.
$$

\noindent Each  strip $cS_\ell^{(i)}(k)$ is naturally decomposed, by the lattice points, into isosceles triangles with base of the triangles parallel to the vector $\db_i$, of length $\sqrt{3}\ell$, and the two equal sides of length $\ell$. 
We denote by $T^{(i)}(\xb^\ell)$ the isosceles triangle belonging to   $cS_\ell^{(i)}$ with vertex in $\xb^\ell$, see Fig. \ref{strips}.

Thanks to these triangulations, we now define, over each strip $cS_\ell^{(i)}(k)$, a piecewise affine function $\hat w_\ell^{(i)}$ that interpolates the lattice values of a given function $w_\ell\in \mathcal{A}_\ell$. The function $\hat w_\ell^{(i)}$ is set to be equal to zero in the complement of $cS_\ell^{(i)}$. We achieve this in two steps. For a given function $w_\ell\in\mathcal{A}_\ell$ and for each $i=1,2,3$, we first define the piecewise affine interpolant of $w_\ell$ over the strips composing $cS_\ell^{(i)}$. That is,
$$
\hat w_\ell^{(i)}\big|_{cS_\ell^{(i)}}: cS_\ell^{(i)}\to\mathbb{R}
$$
is a piecewise affine function with values
$$
\hat w_\ell^{(i)}\big|_{cS_\ell^{(i)}}(\xb^\ell)=w_\ell(\xb^\ell)
$$
for every $\xb^\ell\in L_1(\ell) \cup L_2(\ell)$. We also set $\hat w_\ell^{(i)}:  \mathbb{R}^2\to \mathbb{R}$ by
\begin{equation}\label{eq:defhw}
\hat w_\ell^{(i)}(\xb)=\left\{
\begin{array}{ll}
\hat w_\ell^{(i)}\big|_{cS_\ell^{(i)}}
(\xb) & \xb\in cS_\ell^{(i)},\\[4pt]
0 &\mbox{otherwise}.
\end{array}
\right.
\end{equation}

\begin{figure}[h]
	\centering
	\def\svgwidth{.9\textwidth}
	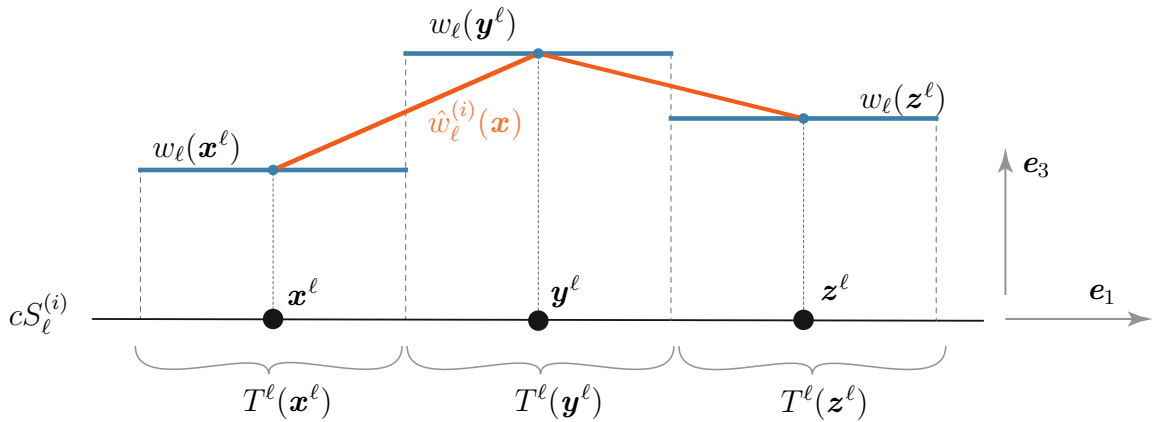
	\caption{ The interpolating function $\hat w^{(i)}_\ell(\xb)$.}
	\label{fig:wl}
\end{figure}

Hence, the point-wise gradient $\nabla \hat w_\ell^{(i)}$ is constant over each triangle $T^{(i)}(\xb^\ell)$;
we denote this constant by $\nabla \hat w_\ell^{(i)}(\xb^\ell)$. Formally, we set
$$
\nabla \hat w_\ell^{(i)}(\xb^\ell):=\nabla \hat w_\ell^{(i)}(\xb)
$$
with $\xb$ any point in $T^{(i)}(\xb^\ell)$.

Among adjacent triangles we may compute the jump of the gradients. 
We denote by
$$
\jump{\nabla \hat w_\ell^{(i)}}_{\pb_j}(\xb^\ell)
$$
the jump of the gradient $\nabla \hat w_\ell^{(i)}$ across triangles, in the union of strips $cS_\ell^{(i)}$, that share a side passing by $\xb^\ell$ and parallel to $\pb_j$. The sign of the jump is computed according to the orientation determined by $\db_i$; that is, it is defined as the value of the gradient  $\nabla \hat w_\ell^{(i)}$ on the triangle on which $\db_i$ is pointing to, minus the value of the gradient over the triangle opposite to the direction of $\db_i$. To become accustomed with the notation introduced we compute \eqref{Z-wedge}$_2$ with $i=1$ and $\xb^\ell\in L_2(\ell)$:
\begin{align}
\Thz_{\pb_1\pb_{3}}[w_\ell](\xb^\ell) &=\frac{2\sqrt{3}}{3\ell}\Big(\hat w_\ell^{(1)}(\xb^\ell+\ell\pb_1-\ell \pb_{3})-\hat w_\ell^{(1)}(\xb^\ell+\ell \pb_{1})\nonumber\\
&\hspace{5cm}+\hat w_\ell^{(1)}(\xb^\ell+\ell \pb_{3})-\hat w_\ell^{(1)}(\xb^\ell)\Big)\nonumber\\
&=\frac{2\sqrt{3}}{3\ell}\Big(\nabla\hat w_\ell^{(1)}(\xb^\ell+\ell\pb_1)\cdot(-\ell \pb_{3})+\nabla\hat w_\ell^{(1)}(\xb^\ell)\cdot (\ell \pb_{3})\Big)\nonumber\\
&=-\frac{2\sqrt{3}}{3}\jump{\nabla\hat w_\ell^{(1)}}_{\pb_1}(\xb^\ell)\cdot \pb_{3},\label{ThetaZ13}
\end{align}
where the first identity holds because $\hat w_\ell^{(1)}$ is affine. 
Similarly, \eqref{Z-wedge}$_1$ with $i=3$ and for $\xb^\ell\in L_2(\ell)$ rewrites as
\begin{align}
\Thz_{\pb_3\pb_{1}}[w_\ell](\xb^\ell) &=\frac{2\sqrt{3}}{3\ell}\Big(\nabla\hat w_\ell^{(1)}(\xb^\ell+\ell\pb_3)\cdot(-\ell \pb_{1})+\nabla\hat w_\ell^{(1)}(\xb^\ell)\cdot (\ell \pb_{1})\Big)\nonumber\\
&=+\frac{2\sqrt{3}}{3}\jump{\nabla\hat w_\ell^{(1)}}_{\pb_3}(\xb^\ell)\cdot \pb_{1}.\label{ThetaZ31}
\end{align}

The $\mathcal{Z}$-- dihedral energy, \eqref{enZ}, can be split into three parts as
\begin{align}
\frac{2\,\mathcal{U}^{\mathcal{Z}}_\ell(w_\ell)}{k^{\mathcal{Z}}}=& \sum_{\xb^\ell\in L_2(\ell)} 
\Big(\Thz_{\pb_1\pb_{3}}[w_\ell](\xb^\ell)\Big)^2+ \Big(\Thz_{\pb_3\pb_{1}}[w_\ell](\xb^\ell)\Big)^2+\nonumber\\
& \sum_{\xb^\ell\in L_2(\ell)} 
\Big(\Thz_{\pb_2\pb_{3}}[w_\ell](\xb^\ell)\Big)^2+ \Big(\Thz_{\pb_3\pb_{2}}[w_\ell](\xb^\ell)\Big)^2+\label{Z-split}\\
& \sum_{\xb^\ell\in L_2(\ell)} 
\Big(\Thz_{\pb_1\pb_{2}}[w_\ell](\xb^\ell)\Big)^2+ \Big(\Thz_{\pb_2\pb_{1}}[w_\ell](\xb^\ell)\Big)^2,\nonumber
\end{align}
where only dihedral bonds contained in the strip $cS_\ell^{(1)}$ are involved in the first line, as it appears also from \eqref{ThetaZ13} and \eqref{ThetaZ31}, while the second and third lines could be written using $\hat w_\ell^{(2)}$ and $\hat w_\ell^{(3)}$, respectively. Thanks to this decomposition we are allowed to study the $\mathcal{Z}$-- dihedral energy only on $cS_\ell^{(1)}$ and then extend the results ``by rotation'' to obtain the equivalent ones on the strips  $cS_\ell^{(2)}$ and $cS_\ell^{(3)}$.

In the next Lemma we establish a bound on the $L^2$-norm of the jumps of $\nabla \hat w_\ell^{(i)}$.

\begin{lem}\label{lem1}
	Let $w_\ell\in\mathcal{A}_\ell$ satisfy the energy bound \eqref{energybound} and let 
	$\hat w_\ell^{(i)}$ be the piecewise affine functions defined in \eqref{eq:defhw}. Then,
	\begin{equation}\label{rescdihedrenrg}
	\sup_\ell\sum_{i=1}^3\ell\,\int_{J(\nabla\hat w^{(i)}_\ell)\cap cS_\ell^{(i)}}|\frac{1}{\ell}\jump{\nabla\hat w^{(i)}_\ell}|^2\, ds<+\infty, 
	\end{equation}
	where  $\jump{\nabla\hat w^{(i)}_\ell}$ and $J(\nabla\hat w_\ell^{(i)})$ denote the jump and the jump set  of $\nabla\hat w_\ell^{(i)}$.
\end{lem}

\proof
We prove the lemma for $i=1$ only, since the other cases can be treated similarly.  Recalling \eqref{ThetaZ13} and \eqref{ThetaZ31} we have that 
\begin{align}
\sum_{\xb^\ell\in L_2(\ell)} &\Big(\Thz_{\pb_1\pb_{3}}[w_\ell](\xb^\ell)\Big)^2+ \Big(\Thz_{\pb_3\pb_{1}}[w_\ell](\xb^\ell)\Big)^2\nonumber\\
&=\frac{4}{3} \sum_{\xb^\ell\in L_2(\ell)} \Big(\jump{\nabla\hat w_\ell^{(1)}}_{\pb_1}(\xb^\ell)\cdot \pb_{3}\Big)^2+\Big(\jump{\nabla\hat w_\ell^{(1)}}_{\pb_3}(\xb^\ell)\cdot \pb_{1}\Big)^2.\label{temp20}
\end{align}
Thanks to the continuity of $\hat w_\ell^{(1)}$ over the strip $cS_\ell^{(1)}$, we find that
$$
\jump{\nabla\hat w_\ell^{(1)}}_{\pb_1}=\jump{\partial_{\pb_1}\hat w_\ell^{(1)} \,{\pb_1}+\partial_{\pb_1^\perp}\hat w_\ell^{(1)}\, {\pb_1^\perp}}_{\pb_1}=\jump{\partial_{\pb_1^\perp}\hat w_\ell^{(1)}}_{\pb_1}\, {\pb_1^\perp},
$$
and therefore
$$
(\jump{\nabla\hat w_\ell^{(1)}}_{\pb_1}\cdot \pb_{3})^2=\frac 34(\jump{\partial_{\pb_1^\perp}\hat w_\ell^{(1)}}_{\pb_1}\, )^2=\frac 34
|\jump{\nabla\hat w_\ell^{(1)}}_{\pb_1}|^2,
$$
since $|{\pb_1^\perp}\cdot\pb_3|=\sqrt{3}/2$.
We may therefore rewrite \eqref{temp20} as
\begin{align*}
\sum_{\xb^\ell\in L_2(\ell)} &\Big(\Thz_{\pb_1\pb_{3}}[w_\ell](\xb^\ell)\Big)^2+ \Big(\Thz_{\pb_3\pb_{1}}[w_\ell](\xb^\ell)\Big)^2\nonumber\\
&=\sum_{\xb^\ell\in L_2(\ell)} |\jump{\nabla\hat w_\ell^{(1)}}_{\pb_1}(\xb^\ell)|^2+|\jump{\nabla\hat w_\ell^{(1)}}_{\pb_3}(\xb^\ell)|^2\\
&=\frac{1}{\ell}\int_{J(\nabla\hat w_\ell^{(1)})\cap cS^{(1)}_\ell}|\jump{\nabla\hat w_\ell^{(1)}}|^2\, ds=\ell\int_{J(\nabla\hat w_\ell^{(1)})\cap cS^{(1)}_\ell}|\frac{1}{\ell}\jump{\nabla\hat w_\ell^{(1)}}|^2\, ds.
\end{align*}
Since $w_\ell\in\mathcal{A}_\ell$ satisfy the energy bound \eqref{energybound}, we have that
$\sup_\ell \mathcal{U}^{\mathcal{Z}}_\ell(w_\ell)<+\infty$ and hence \eqref{rescdihedrenrg} follows by recalling \eqref{Z-split}. \QED

We now prove that $\hat w^{(i)}_\ell$ is $H^1$-bounded over the region $cS_\ell^{(i)}$.

\begin{lem}\label{lem2}
	Let $\hat w^{(i)}_\ell$ be as in Lemma~\ref{lem1}. Then,
	\begin{equation}\label{H1bound}
	\sup_\ell\,\|\hat w^{(i)}_\ell\|_{H^1(cS_\ell^{(i)})}<+\infty.
	\end{equation}
\end{lem}

\proof
Again we prove the lemma for $i=1$.
This proof is more transparent if we adopt a  notation different from that used so far.   
For given $k$, we denote the triangles on the strip $cS_\ell^{(1)}(k)$ by $T_j$ (instead of $T^{(1)}(\xb^\ell)$), with $j\in\mathbb{Z}$ increasing in the direction $\db_1$, and we write $\nabla\hat w^{(1)}_\ell(T_j)$ to indicate the constant value taken by $\nabla\hat w^{(1)}_\ell$ over the triangle $T_j$. Notice that the following identity 
$$
\nabla\hat w^{(1)}_\ell(T_j)=\sum_{m=-\infty}^j\left(\nabla\hat w^{(1)}_\ell(T_{m})-\nabla\hat w^{(1)}_\ell(T_{m-1})\right)
$$
holds because on the right we have a telescoping sum and because $\hat w^{(1)}_\ell$ vanishes outside of $\Omega$. 
Hence,
\begin{align*}
|\nabla\hat w^{(1)}_\ell(T_j)|&\le\sum_{m=-\infty}^\infty|\left(\nabla\hat w^{(1)}_\ell(T_{m})-\nabla\hat w^{(1)}_\ell(T_{m-1})\right)|\\
&=\int_{J(\nabla\hat w^{(1)}_\ell)\cap cS_\ell^{(1)}(k)}|\frac{1}{\ell}\jump{\nabla\hat w^{(1)}_\ell}|\, ds,
\end{align*}
and by applying Jensen inequality we find
\begin{align*}
|\nabla\hat w^{(1)}_\ell(T_j)|^2&\le|J(\nabla\hat w^{(1)}_\ell)\cap cS_\ell^{(1)}(k)|\int_{J(\nabla\hat w^{(1)}_\ell)\cap cS_\ell^{(1)}(k)}|\frac{1}{\ell}\jump{\nabla\hat w^{(1)}_\ell}|^2\, ds\\
&\le C \int_{J(\nabla\hat w^{(1)}_\ell)\cap cS_\ell^{(1)}(k)}|\frac{1}{\ell}\jump{\nabla\hat w^{(1)}_\ell}|^2\, ds.
\end{align*}
The last inequality follows because $|J(\nabla\hat w^{(1)}_\ell)\cap cS_\ell^{(1)}(k)|\le c\ell\,(\mbox{diam}\Omega/\ell)$.
Thence, multiplying by $|T_j|$ on both sides and summing over $j$ we get
\begin{align*}
\|\nabla\hat w^{(1)}_\ell\|_{L^2(cS_\ell^{(1)}(k))}^2&=\sum_{T_j\cap\Omega\ne\emptyset}|\nabla\hat w^{(1)}_\ell(T_j)|^2|T_j|\\
&\le c\sum_{T_j\cap\Omega\ne\emptyset}|T_j| \int_{J(\nabla\hat w^{(1)}_\ell)\cap cS_\ell^{(1)}(k)}|\frac{1}{\ell}\jump{\nabla\hat w^{(1)}_\ell}|^2\, ds\\
&\le c\,\mbox{diam }\Omega\,\ell \,\int_{J(\nabla\hat w^{(1)}_\ell)\cap cS_\ell^{(1)}(k)}|\frac{1}{\ell}\jump{\nabla\hat w^{(1)}_\ell}|^2\, ds.
\end{align*}
Hence, summing over $k$ and taking Lemma~\ref{lem1} into account, we have 
$$
\|\nabla\hat w^{(1)}_\ell\|_{L^2(cS_\ell^{(1)})}<+\infty.
$$
With Poincar\'e inequality we deduce \eqref{H1bound}. \QED

We are now in a position to prove \eqref{wlw}  of Theorem \ref{thm-wstrong_0};  the regularity of the limit function will be proved later.

\begin{thm}\label{thm-wstrong}
	Let $w_\ell\in\mathcal{A}_\ell$ satisfy the energy bound \eqref{energybound}. Then, there exists $w\in L^2(\Ro^2)$ 
	equal to zero almost everywhere outside of $\Omega$ such that, up to a subsequence,
	\begin{equation}\label{strong_wlt5}
	w_\ell\to w \quad \mbox{in }L^2(\mathbb{R}^2).
	\end{equation}
\end{thm}

\proof
Note that $w_\ell$ is uniformly bounded in $L^2(\mathbb{R}^2)$. Indeed,
$$
w_\ell(\xb)=\sum_{\xb^\ell\in L_1(\ell)\cup L_2(\ell)}w_\ell(\xb^\ell)\ \chi_{T^\ell(\xb^\ell)}(\xb)
=\sum_{\xb^\ell\in L_1(\ell)\cup L_2(\ell)}\hat w^{(1)}_\ell(\xb^\ell)\ \chi_{T^\ell(\xb^\ell)}(\xb),
$$
hence
\begin{align}\label{boundwl_2}
\int_{\mathbb{R}^2}|w_\ell|^2\, d\xb&\le \sum_{\xb^\ell\in L_1(\ell)\cup L_2(\ell)}|\hat w^{(1)}_\ell(\xb^\ell)|^2|T^\ell(\xb^\ell)(\xb)|\nonumber \\
&\le c\sum_{\xb^\ell\in L_1(\ell)\cup L_2(\ell)}\int_{T^{(1)}(\xb^\ell)}|\hat w_\ell^{(1)}(\xb)|^2+|\nabla\hat w_\ell^{(1)}(\xb)|^2 \ell^2\,d\xb
\nonumber \\
&\le c \|\hat w_\ell^{(1)}\|^2_{H^1(cS_\ell^{(1)})}\le C
\end{align}
by Lemma~\ref{lem2}. 
Then, there exists $w\in L^2(\Ro^2)$ 
equal to zero almost everywhere outside of $\Omega$ such that, up to a subsequence,
\begin{equation}\label{weak_wl}
w_\ell\rightharpoonup w \quad \mbox{in }L^2(\mathbb{R}^2).
\end{equation}
We now prove that the convergence is in fact strong.

Let $\hb=h^1\db_1+h^2\db_2 \in\mathbb{R}^2$ and write 
\begin{equation}\label{whd00}
\begin{aligned}
\int_{\mathbb{R}^2}|w_\ell(\xb+\hb)-w_\ell(\xb)|^2\,d\xb&\le C \int_{\mathbb{R}^2}|w_\ell(\xb+h^1\db_1+h^2\db_2)- w_\ell(\xb+h^1\db_1)|^2\,d\xb\\
&+C\int_{\mathbb{R}^2}|w_\ell(\xb+h^1\db_1)-w_\ell(\xb)|^2\,d\xb.
\end{aligned}
\end{equation}
Let us consider the second term on the right hand side (the first can be handled similarly). For each $\xb$ and each $h^1$ there exist two lattice poins
$\yb^\ell$ and $\zb^\ell$ such that
\begin{equation}\label{xlyl0}
\xb \in T^\ell(\yb^\ell), \quad \mbox{and}\quad \xb+h^1\db_1 \in T^\ell(\zb^\ell).
\end{equation}
With the notation introduced in \eqref{eqnew} we may also write
\begin{equation}\label{xlyl}
\begin{aligned}
\yb^\ell & = n_y\ell\db_1+k\ell\db_2+m_y\ell\pb, \\
\zb^\ell & = n_z\ell\db_1+k\ell\db_2+m_z\ell\pb,
\end{aligned}
\end{equation}
with $n_y,n_z,k\in\mathbb{Z}$ and $m_y,m_z\in\{0,1\}$.

Without loss of generality we assume that $h^1>0$, which implies that $n_y\ge n_x$.

We consider a ``monotonic'' path from $\yb^\ell$ to $\zb^\ell$ through the lattice points of the strip $S^{(1)}_\ell(k)$ and label these points by means of the index $j=0,\dots,j_f$, so that 
$$
\xb(0)=\yb^\ell \quad \mbox{and} \quad \xb(j_f)=\zb^\ell.
$$
Then, the number of sides of the path are $2(n_z-n_y)$ or $2(n_z-n_y)\pm1$ according to whether the initial and the final points belong to the same Bravais lattice, {\it i.e.}, $m_z=m_y$.

With this notation we have:
\begin{align}
| w_\ell(\xb+h^1&\db_1)- w_\ell(\xb)|^2
=| w_\ell(\yb^\ell)- w_\ell(\xb^\ell)|^2
=|\hat w^{(1)}_\ell(\yb^\ell)-\hat w^{(1)}_\ell(\xb^\ell)|^2\nonumber\\
&=|\sum_{j=1}^{j_f}\hat w^{(1)}_\ell(\xb(j))-\hat w^{(1)}_\ell(\xb(j-1))|^2\nonumber\\
&\le |2(n_z-n_y)+1|\sum_{j=1}^{j_f}|\hat w^{(1)}_\ell(\xb(j))-\hat w^{(1)}_\ell(\xb(j-1))|^2\nonumber\\
&\le|2(n_z-n_y)+1|\sum_{j\in \mathbb{Z}}\left|\nabla\hat w^{(1)}_\ell\xb(j))\right|^2\ell^2\nonumber\\
&\le c |2(n_z-n_y)+1|\int_{cS_\ell^{(1)}(k)}|\nabla\hat w^{(1)}_\ell|^2\,d\xb.\label{boh1}
\end{align}
We now estimate $2(n_z-n_y)+1$.

\begin{figure}[h]
	\centering
	\def\svgwidth{.3\textwidth}
	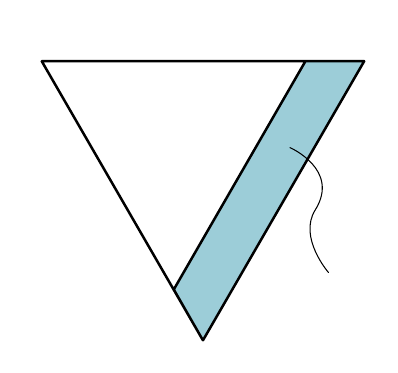
	\caption{The strip $S^h(\xb^\ell)$}\label{stripsm}
\end{figure}

We first look at the case $h^1>\ell$. From the obvious inequality 
$$
(2(n_z-n_y)-1)\ell\frac{\sqrt{3}}{2}\le|(\zb^\ell-\yb^\ell)\cdot \frac{\db_1}{|\db_1|} |\le|\zb^\ell-\yb^\ell |,
$$
we find
$$
2(n_z-n_y)+1\le\frac{2\sqrt{3}}{3\ell}|\zb^\ell-\yb^\ell |+2\le \frac{2\sqrt{3}}{3\ell}|\zb^\ell-\yb^\ell |+2\frac{h^1}\ell,
$$
and noting that $|\zb^\ell-\yb^\ell |\le |h^1\db_1| + \ell=\sqrt{3}h^1+\ell$, we deduce that
\begin{equation}\label{nynx}
2(n_z-n_y)+1\le
\frac{2\sqrt{3}}{3\ell}(\sqrt{3}h^1+\ell)+2\frac{h^1}\ell
\le c \frac{1}{\ell}|\bf h|.
\end{equation}
Then, by integrating  \eqref{boh1} over the union of the triangles $T^{\ell}(\xb^\ell)$, with $\xb^\ell\in S^{(1)}_\ell(k)\cap\Omega$:
$$
A(k):= \bigcup_{\xb^\ell\in S^{(1)}_\ell(k)\cap\Omega}T^{\ell}(\xb^\ell),
$$
we get
$$
\int_{A(k)}|w_\ell(\xb+h^1\db_1)-w_\ell(\xb)|^2\,d\xb\le c \frac{|\hb|}{\ell} \mbox{diam}(\Omega)\, \frac \ell{2} \int_{cS_\ell^{(1)}(k)}|\nabla\hat w^{(1)}_\ell|^2\,d\xb.
$$
Hence, by summing over $k$ and applying Lemma \ref{lem2}, we have that
\begin{equation}\label{bound_1}
\int_{\mathbb{R}^2}| w_\ell(\xb+h^1\db_1)- w_\ell(\xb)|^2\,d\xb\le C |\hb|.
\end{equation}
We now look at the case $h^1<\ell$. We have
\begin{align}
\int_{\mathbb{R}^2}| w_\ell(\xb+&h^1\db_1)-w_\ell(\xb)|^2\,d\xb\nonumber\\
&=
\sum_{\xb^\ell\in L_1(\ell)\cup L_2(\ell)}\int_{T^\ell(\xb^\ell)}| w_\ell(\xb+h^1\db_1)-w_\ell(\xb)|^2\,d\xb\nonumber\\
&=\sum_{\xb^\ell\in L_1(\ell)\cup L_2(\ell)}\int_{T^\ell(\xb^\ell)\cap S^h(\xb^\ell)}|w_\ell(\xb+h^1\db_1)- w_\ell(\xb)|^2\,d\xb,\label{boh2}
\end{align}
where $S^h(\xb^\ell)$ is a strip contained in $T^\ell(\xb^\ell)$ of width 
$|h_1\db_1|$, in the direction $\db_1$, and sides parallel to $\db_2$ or $\db_3$, see Figure~\ref{stripsm}.
The third equality follows since the difference $w_\ell(\xb+h^1\db_1)- w_\ell(\xb)$ vanishes for all $\xb \in T^\ell(\xb^\ell)\setminus S^h(\xb^\ell)$. 
In this case for every $\xb \in S^{h}(\xb^\ell)$ the point $(\xb+h^1\db_1)$ has to belong to the next or after next neighbor triangle. With the notation above, one calculates that 
$$
2(n_z-n_y)+1\le 3,
$$
and from \eqref{boh1} and \eqref{boh2}, we deduce  
\begin{align}\label{bound_2}
\int_{\mathbb{R}^2}|w_\ell(\xb+h^1&\db_1)- w_\ell(\xb)|^2\,d\xb\nonumber\\
&\le
c\sum_{\xb^\ell\in (L_1(\ell)\cup L_2(\ell))\cap\Omega}\int_{T^\ell(\xb^\ell)\cap S^h(\xb^\ell)}\Big(\int_{cS_\ell^{(1)}(k)}|\nabla\hat w^{(1)}_\ell|^2\,d\xb\Big)\,d\xb\nonumber\\
&\le c\sum_{\xb^\ell\in (L_1(\ell)\cup L_2(\ell))\cap\Omega}|\hb|\ell\int_{cS_\ell^{(1)}(k)}|\nabla\hat w^{(1)}_\ell|^2\,d\xb\nonumber\\
&=c\sum_{k\in\mathbb{Z}}\sum_{\xb^\ell\in cS_\ell^{(1)}(k)\cap\Omega}|\hb|\ell\int_{cS_\ell^{(1)}(k)}|\nabla\hat w^{(1)}_\ell|^2\,d\xb\nonumber\\
&=c\sum_{k\in\mathbb{Z}}\mbox{diam}(\Omega)|\hb|\int_{cS_\ell^{(1)}(k)}|\nabla\hat w^{(1)}_\ell|^2\,d\xb\nonumber\\
&\le c |\hb|\int_{cS_\ell^{(1)}}|\nabla\hat w^{(1)}_\ell|^2\,d\xb\le C |\hb|,
\end{align}
where we have taken into account that the number of lattice points in  $S_\ell^{(1)}(k)\cap\Omega$  is of order $\mbox{diam}(\Omega)/\ell$.

From \eqref{bound_1} and \eqref{bound_2} we get that
$$
\int_{\mathbb{R}^2}|w_\ell(\xb+h^1\db_1)- w_\ell(\xb)|^2\,d\xb\le C |\hb|
$$
uniformly in $\ell$.
Since the first term on the right hand side of \eqref{whd00} can be studied in exactly the same way, we have that
for every $\hb\in\Ro^2$
\begin{equation}\label{bound_3}
\int_{\mathbb{R}^2}|w_\ell(\xb+\hb)- w_\ell(\xb)|^2\,d\xb\le C |\hb|
\end{equation}
uniformly in $\ell$.
Then, by \eqref{bound_3}, \eqref{boundwl_2},  and Riesz-Kolmogorov's theorem it follows that 
$w_\ell$ has a strongly convergent subsequence. Thus the convergence stated in \eqref{weak_wl} is strong. \QED

We end this section with two lemmas that address the convergence of $\hat w^{(i)}_\ell$ and its derivatives.
\begin{lem}\label{lemwhw0}
	Let $\hat w^{(i)}_\ell$ be as in Lemma~\ref{lem1} and $w$ be as in Theorem \ref{thm-wstrong}. Then,
	$w\in H^1(\Ro^2)$ and  there exists a subsequence, not relabeled,  such that
	\begin{equation}\label{derwdi}
	\hat w^{(i)}_\ell\rightharpoonup \frac 13 w, \qquad \partial_{\db_i}\hat w^{(i)}_\ell \rightharpoonup\frac{1}{3}\partial_{\db_i}w\quad \mbox{in }L^2(\mathbb{R}^2),
	\end{equation}
	for $i=1,2,3.$
\end{lem}

\proof 
From the definition \eqref{eq:defhw} of $\hat w^{(i)}_\ell$, we have that $\partial_{\db_i}\hat w^{(i)}_\ell=0 \ \mbox{in} \ \mathbb{R}^2\setminus cS_\ell^{(i)}$. Hence, by Lemma \ref{lem2} it follows that 
\begin{equation}\label{boundwi_1}
\|\hat w^{(i)}_\ell\|_{L^2(\mathbb{R}^2)}+\| \partial_{\db_1}\hat w^{(i)}_\ell\|_{L^2(\mathbb{R}^2)}<+\infty.
\end{equation}
Up to a subsequence, we have that
\begin{equation}\label{boundwi_2}
\hat w^{(i)}_\ell\rightharpoonup \hat w^{(i)}, \qquad \partial_{\db_i}\hat w^{(i)}_\ell \rightharpoonup\partial_{\db_i}\hat w^{(i)}\quad \mbox{in }L^2(\mathbb{R}^2),
\end{equation}
for some $\hat w^{(i)}\in L^2(\Ro^2)$ with $\partial_{\db_i}\hat w^{(i)}\in L^2(\Ro^2)$.

Let $\psi\in C_0^\infty(\mathbb{R}^2)$. Then, 
\begin{align}\label{whw}
\int_{\mathbb{R}^2}\hat w^{(i)}_\ell\psi\,d\xb=\int_{\mathbb{R}^2}\hat w^{(i)}_\ell\chi_{cS^{(i)}_\ell}\psi\,d\xb=\int_{\mathbb{R}^2}w_\ell\chi_{cS^{(i)}_\ell}\psi\,d\xb+\int_{cS^{(i)}_\ell}(\hat w^{(i)}_\ell- w_\ell)\psi\,d\xb.
\end{align}
The last integral on the right hand side tends to zero, since
$$
|\int_{cS^{(i)}_\ell}(\hat w^{(i)}_\ell-w_\ell)\psi\,d\xb|\le \int_{cS^{(i)}_\ell}|\nabla\hat w^{(i)}_\ell|\ell|\psi|\,d\xb\le \ell \|\nabla\hat w^{(i)}_\ell\|_{L^2(cS^{(i)}_\ell)}\|\psi\|_{L
	^2(\Ro^2)}.
$$
By Theorem~\ref{thm-wstrong} and taking into account that $\chi_{cS^{(i)}_\ell}\overset{*}\rightharpoonup 1/3$ in $L^\infty(\Ro^2)$, passing to the limit in \eqref{whw} yields that
$$
\int_{\mathbb{R}^2} \hat w^{(i)}\psi\,d\xb=\int_{\mathbb{R}^2}\frac 13 w\psi\,d\xb
$$
from which we deduce that $\hat w^{(i)}=\frac 13 w$, for $i=1,2,3.$
But,  by \eqref{boundwi_2},
$\partial_{\db_i} w\in L^2(\mathbb{R}^2)$ for $i=1,2,3$
and, since $w\in L^2(\mathbb{R}^2)$, we have $w\in H^1(\mathbb{R}^2)$.
\QED

In the previous lemma we have deduced the weak limit of $\partial_{\db_i}\hat w^{(i)}_\ell$.
We notice that, by the definition of $\hat w^{(i)}_\ell$, the pointwise derivative of  $\hat w^{(i)}_\ell$ in the direction $\db_i$ coincides with the distributional derivative in the same direction. This is not the case for other directional  derivatives, because the distributional gradient of $\hat w^{(i)}_\ell$ is singular at $\partial cS^{(i)}_\ell$ due to the discontinuity of $\hat w^{(i)}_\ell$. Below we denote by $\gb^{(i)}_\ell$ the absolutely continuous  part, with respect to the 2-dimensional Lebesgue measure, of the distributional gradient of $\hat w^{(i)}_\ell$ and give a characterization of its limit in the next theorem.

\begin{thm}\label{gi}
	Let $\hat w^{(i)}_\ell$ be as in Lemma~\ref{lem1} and let 
	\begin{equation}\label{gli}
	\gb^{(i)}_\ell(\xb)=\left\{
	\begin{array}{ll}
	\nabla \hat w_\ell^{(i)}
	(\xb) & \xb\in cS_\ell^{(i)}, \\[2pt]
	\mathbf{0}&\mbox{otherwise}.
	\end{array}
	\right.
	\end{equation}
	Then, up to subsequences, 
	\begin{equation}\label{weakgli}
	\gb^{(i)}_\ell\rightharpoonup \gb^{(i)} \quad\mbox{in }L^2(\Ro^2;\Ro^2),
	\end{equation}
	where, with $w$ as in Lemma \ref{lemwhw0},
	\begin{equation}\label{girep}
	\gb^{(1)}=\frac 13\nabla w+\gamma \pb_2,\quad \gb^{(2)}=\frac 13\nabla w+\gamma \pb_1,\quad \gb^{(3)}=\frac 13\nabla w+\gamma \pb_3,
	\end{equation}
	with $\gamma\in L^2(\Ro^2)$ and equal to zero almost everywhere outside of $\Omega$.
\end{thm}

\proof
By Lemma \ref{lem2}, the weak convergence stated in \eqref{weakgli} holds for $\gb^{(i)} \in L^2(\Ro^2;\Ro^2)$ and for a subsequence.
By  definition \eqref{gli}, the equalities
$$
\gb^{(i)}_\ell\cdot \frac{\db_i}{|\db_i|}=\partial_{\db_i}\hat w^{(i)}_\ell \quad \mbox{for }i=1,2,3,
$$
hold true almost everywhere in $\Ro^2$ and not just in $cS^{(i)}_\ell$, because $\partial_{\db_i}\hat w^{(i)}_\ell=0$ in $\Ro^2 \setminus cS^{(i)}_\ell$. Then, from  \eqref{derwdi}  and by passing to the limit we find that
$$
\gb^{(i)}\cdot \frac{\db_i}{|\db_i|}=\frac{1}{3}\partial_{\db_i} w \quad \mbox{for }i=1,2,3.
$$
Since $\db_1\cdot \pb_2= \db_2\cdot \pb_1= \db_3\cdot \pb_3=0$, we can write
\begin{equation}\label{chrctgi}
\gb^{(1)}=\frac 13\nabla w+\gamma^{(1)} \pb_2,\quad \gb^{(2)}=\frac 13\nabla w+\gamma^{(2)} \pb_1,\quad \gb^{(3)}=\frac 13\nabla w+\gamma^{(3)} \pb_3
\end{equation}
with $\gamma^{(i)}\in L^2(\Ro^2)$.

We now show that 
\begin{equation}\label{g1g3}
\gb^{(1)}\cdot \pb_3=\gb^{(2)}\cdot \pb_3, \quad \gb^{(1)}\cdot \pb_1=\gb^{(3)}\cdot \pb_1,\quad \gb^{(3)}\cdot \pb_2=\gb^{(2)}\cdot \pb_2,
\end{equation}
hold almost everywhere in $\Ro^2$. We limit ourselves to the proof of $\gb^{(1)}\cdot \pb_1=\gb^{(3)}\cdot \pb_1$, since the other equalities can be proved similarly.

\begin{figure}[h]
	\centering
	\def\svgwidth{1\textwidth}
	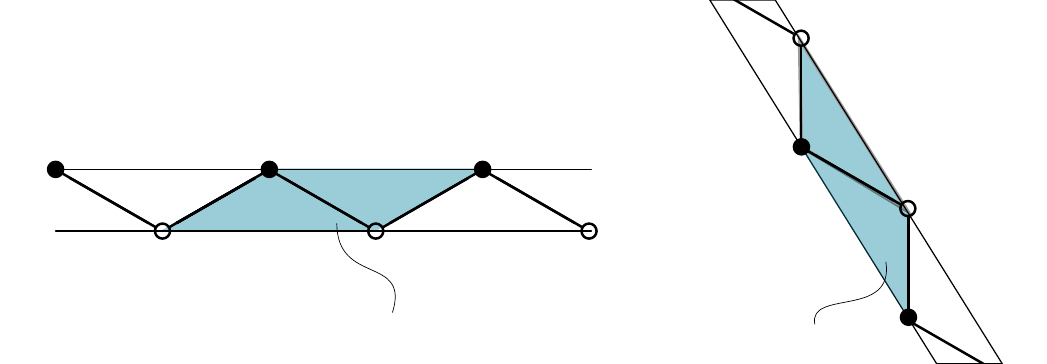
	\caption{The regions $\Pc^{(1)}(\xb^\ell)$ and $\Pc^{(3)}(\xb^\ell)$.}
	\label{fig:triangles}
\end{figure}

Let $\xb^\ell \in L_2(\ell)$ and define
\begin{align*}
\Pc^{(1)}(\xb^\ell)&:=T^{(1)}(\xb^\ell)\cup T^{(1)}(\xb^\ell+\ell\pb_1),\\
\Pc^{(3)}(\xb^\ell)&:=T^{(3)}(\xb^\ell)\cup T^{(3)}(\xb^\ell+\ell\pb_1),
\end{align*}
see Fig. \ref{fig:triangles}.

Since the triangles $T^{(1)}(\xb^\ell)$ and $T^{(1)}(\xb^\ell+\ell\pb_1)$ have a common side
parallel to $\pb_1$, it follows that $\partial_{\pb_1}\hat w^{(1)}_\ell$ is constant on $\Pc^{(1)}(\xb^\ell)$.
Similarly, $\partial_{\pb_1}\hat w^{(3)}_\ell$ is constant on $\Pc^{(3)}(\xb^\ell)$.
Furthermore, on the segment joining $\xb^\ell$ to $\xb^\ell+\ell \pb_1$ we have  $\partial_{\pb_1}\hat w^{(1)}_\ell=\partial_{\pb_1}\hat w^{(3)}_\ell$.
Hence, we have that
\begin{equation}\label{P13}
\int_{\Pc^{(1)}(\xb^\ell)}\partial_{\pb_1}\hat w^{(1)}_\ell\,d\xb=\int_{\Pc^{(3)}(\xb^\ell)}\partial_{\pb_1}\hat w^{(3)}_\ell\,d\xb
\end{equation}
for all lattice point $\xb^\ell\in L_2(\ell)$. 
Let $B\subset \Ro^2$ be an open subset, and let $B_\ell:=\{x\in B: \mbox{ dist}(x,\partial B)\ge \ell\}$. 
If  the segment joining $\xb^\ell$ to $\xb^\ell+\ell \pb_1$ is contained in $B_\ell$
then  $\Pc^{(1)}(\xb^\ell)\cup\Pc^{(3)}(\xb^\ell)\subset B$ and hence, by \eqref{P13},
$$
\int_{\Pc^{(1)}(\xb^\ell)\cap B}\partial_{\pb_1}\hat w^{(1)}_\ell\,d\xb=\int_{\Pc^{(3)}(\xb^\ell)\cap B}\partial_{\pb_1}\hat w^{(3)}_\ell\,d\xb.
$$
Therefore,
\begin{align*}
|\int_{B}\partial_{\pb_1}\hat w^{(1)}_\ell-\partial_{\pb_1}\hat w^{(3)}_\ell\,d\xb| & \le \int_{B\setminus B_\ell}|\partial_{\pb_1}\hat w^{(1)}_\ell-\partial_{\pb_1}\hat w^{(3)}_\ell|\,d\xb\\
&\le |B\setminus B_\ell|^{\frac 12} \|\partial_{\pb_1}\hat w^{(1)}_\ell-\partial_{\pb_1}\hat w^{(3)}_\ell\|_{L^2(B)}
\le C\sqrt{\ell},
\end{align*}
where to obtain the last bound we used \eqref{H1bound}.
By the definition of $\gb^{(i)}_\ell$ it follows that
$$
|\int_{B}\gb^{(1)}_\ell\cdot \pb_1-\gb^{(3)}_\ell\cdot\pb_1\,d\xb| 
\le C\sqrt{\ell},
$$
and by passing to the limit we find
$$
\int_{B}\gb^{(1)}\cdot \pb_1-\gb^{(3)}\cdot\pb_1\,d\xb=0.
$$
Since this identity holds for every open set $B$ we deduce that $\gb^{(1)}\cdot \pb_1=\gb^{(3)}\cdot \pb_1$ almost everywhere in $\Ro^2$.  
This proves \eqref{g1g3}.

From \eqref{chrctgi} and \eqref{g1g3} we find that
$$
\begin{array}{l}
\gb^{(1)}\cdot \pb_3-\gb^{(2)}\cdot \pb_3=\frac{1}{2}(\gamma^{(2)}-\gamma^{(1)})=0,\\ 
\gb^{(1)}\cdot \pb_1-\gb^{(3)}\cdot \pb_1=\frac{1}{2}(\gamma^{(3)}-\gamma^{(1)})=0,\\
\gb^{(3)}\cdot \pb_2-\gb^{(2)}\cdot \pb_2=\frac{1}{2}(\gamma^{(3)}-\gamma^{(2)})=0,
\end{array}
$$
and this implies that
$
\gamma^{(1)}=\gamma^{(2)}=\gamma^{(3)}=:\gamma.
$
That $\gamma$ is equal to zero almost everywhere outside of $\Omega$ it follows since $\nabla \hat w^{(i)}_\ell$ is equal to zero in that region.
\QED

The results obtained so far hold for whatever $k^{\mathcal{C}}$ and $\tau_0$. In the next theorem we show that we can further specify $\gb^{(i)}$ if either one of these two constants is different from zero.

\begin{thm}\label{gamma}
	Let $w_\ell\in\mathcal{A}_\ell$ and $\hat w^{(i)}_\ell$ be as in Lemma~\ref{lem1}, and let $\gamma$ be as in Theorem \ref{gi}.
	If either $k^{\mathcal{C}}\ne 0$ or $\tau_0\ne 0$, then $\gamma=0$ almost everywhere in $\Ro^2$.
\end{thm}

\proof 
Let us first consider the case $k^{\mathcal{C}}\ne 0$. Then, by assumption $k^{\mathcal{C}}> 0$, and since 
$w_\ell$ satisfies the energy bound \eqref{energybound} we find, among other things, that
\begin{equation}\label{boundp2}
\sum_{\xb^\ell\in L_2(\ell)} \Big(\Thc_{\pb_2^+}[w_\ell](\xb^\ell)\Big)^2<+\infty.
\end{equation}
Set
$$
\qcl(\xb):=\sum_{\xb^\ell\in L_2(\ell)} \Thc_{\pb_2^+}[w_\ell](\xb^\ell)\chi_{T^\ell(\xb^\ell)}(\xb).
$$
Then,  $\qcl \to 0$ in $L^2(\Ro^2)$. Indeed,  
since the area of $T^\ell(\xb^\ell)$ is equal to $3\sqrt{3}\ell^2/4$, we have
$$
\int_{\Ro^2} (\qcl )^2\,d\xb= \frac{3\sqrt{3}\ell^2}4 \sum_{\xb^\ell\in L_2(\ell)} \Big(\Thc_{\pb_2^+}[w_\ell](\xb^\ell)\Big)^2,
$$
which converges to zero, as $\ell$ goes to zero, by \eqref{boundp2}.

Let $B\subset \Ro^2$ be any bounded and open set. Then,
\begin{align*}
0=\lim_{\ell\to 0}\int_B \qcl\,d\xb=\lim_{\ell\to 0} \frac{3\sqrt{3}\ell^2}4 \sum_{\xb^\ell\in L_2(\ell)\cap B} \Thc_{\pb_2^+}[w_\ell](\xb^\ell)
\end{align*}
and since, by \eqref{C-wedge}, we have that
\begin{align*}
\Thc_{\pb_2^+}[w_\ell](\xb^\ell)&=\frac{2\sqrt{3}}{3\ell}\big(w_\ell(\xb^\ell+\ell\pb_2-\ell \pb_{1}) -w_\ell(\xb^\ell+\ell \pb_{2})\\
&\hspace{1,5cm}-w_\ell(\xb^\ell+\ell \pb_{2})+w_\ell(\xb^\ell)+
w_\ell(\xb^\ell)-w_\ell(\xb^\ell+\ell \pb_{3})\big)\\
&=\frac{2\sqrt{3}}{3}(-\pb_1\cdot \nabla\hat w_\ell^{(1)}(\xb^\ell+\ell\pb_2) -\pb_2\cdot \nabla\hat w_\ell^{(2)}(\xb^\ell)\\
&\hspace{1,5cm}-\pb_3\cdot \nabla\hat w_\ell^{(1)}(\xb^\ell)),
\end{align*}
we can write 
\begin{align*}
0&=\lim_{\ell\to 0} \frac{3\ell^2}2 \sum_{\xb^\ell\in L_2(\ell)\cap B} (-\pb_1\cdot \nabla\hat w_\ell^{(1)}(\xb^\ell+\ell\pb_2) -\pb_2\cdot \nabla\hat w_\ell^{(2)}(\xb^\ell)\\
&\hspace{3,7cm}-
\pb_3\cdot \nabla\hat w_\ell^{(1)}(\xb^\ell))\\
&=\lim_{\ell\to 0} \sqrt{3} \sum_{\xb^\ell\in L_2(\ell)\cap B} \Big(-\int_{T^{(1)}(\xb^\ell+\ell\pb_2)\cup T^{(1)}(\xb^\ell+\ell\pb_2-\ell\pb_1)}\pb_1\cdot \nabla\hat w_\ell^{(1)}(\xb)\,d\xb \\
&\hspace{3,7cm}-\int_{T^{(2)}(\xb^\ell)\cup T^{(2)}(\xb^\ell+\ell\pb_2)}\pb_2\cdot \nabla\hat w_\ell^{(2)}(\xb)\,d\xb\\
&\hspace{3,7cm}-\int_{T^{(1)}(\xb^\ell)\cup T^{(1)}(\xb^\ell+\ell\pb_3)}
\pb_3\cdot \nabla\hat w_\ell^{(1)}(\xb)\,d\xb \Big),
\end{align*}
since, for instance, $\pb_3\cdot \nabla\hat w_\ell^{(1)}$ is constant on $T^{(1)}(\xb^\ell)\cup T^{(1)}(\xb^\ell+\ell\pb_3)$.
Recalling the definition of $\gb^{(i)}_\ell$, we may rewrite the previous equality as
$$
0=\lim_{\ell\to 0} \int_{B}\pb_1\cdot \gb^{(1)}_\ell(\xb)+\pb_2\cdot \gb^{(2)}_\ell(\xb)+
\pb_3\cdot \gb^{(1)}_\ell(\xb)\,d\xb ,
$$
which, by \eqref{weakgli}, implies that
$$
\int_{B}\pb_1\cdot \gb^{(1)}(\xb)+\pb_2\cdot \gb^{(2)}(\xb)+
\pb_3\cdot \gb^{(1)}(\xb)\,d\xb =0.
$$
From the arbitrariness of the set $B$ we find that 
$$
\pb_1\cdot \gb^{(1)}+\pb_2\cdot \gb^{(2)}+
\pb_3\cdot \gb^{(1)}=0,
$$
almost everywhere in $\Ro^2$.
This is equivalent, by \eqref{girep}, to
$$
\frac 13 \nabla w\cdot (\pb_1+\pb_2+\pb_3)-\frac 32 \gamma=0,
$$
which implies that $\gamma=0$, since $\pb_1+\pb_2+\pb_3=\boldsymbol 0$.

The proof for the case $\tau_0\ne 0$ is similar; hereafter we only sketch it.
For
$$
\qsl(\xb):=\sum_{\xb^\ell\in L_2(\ell)} \Ths_{2}[w_\ell](\xb^\ell)\chi_{T^\ell(\xb^\ell)}(\xb),
$$
we have that 
$$
\lim_{\ell\to 0}\int_B \qsl\,d\xb=0
$$
for any bounded open set $B$ of $\Ro^2$. Hence, thanks to \eqref{s_angles}, we 
may write
\begin{align*}
\Ths_{2}[w_\ell](\xb^\ell)
&=\frac{\sqrt{3\sqrt{3}}}{3}(\pb_1\cdot \nabla\hat w_\ell^{(1)}(\xb^\ell) +\pb_2\cdot \nabla\hat w_\ell^{(2)}(\xb^\ell)
-\pb_3\cdot \nabla\hat w_\ell^{(1)}(\xb^\ell)),
\end{align*}
which let us  arrive at
$$
\lim_{\ell\to 0} \int_{B}\pb_1\cdot \gb^{(1)}_\ell(\xb)+\pb_2\cdot \gb^{(2)}_\ell(\xb)+
\pb_3\cdot \gb^{(1)}_\ell(\xb)\,d\xb=0.
$$
This identity implies, as shown above, that $\gamma=0$ almost everywhere in $\Ro^2$.
\QED

\section{Lower bounds and proof of Theorem \ref{thm-wstrong_0}}\label{sec:Zdihedr}

\begin{figure}[h]
	\centering
	\def\svgwidth{1\textwidth}
	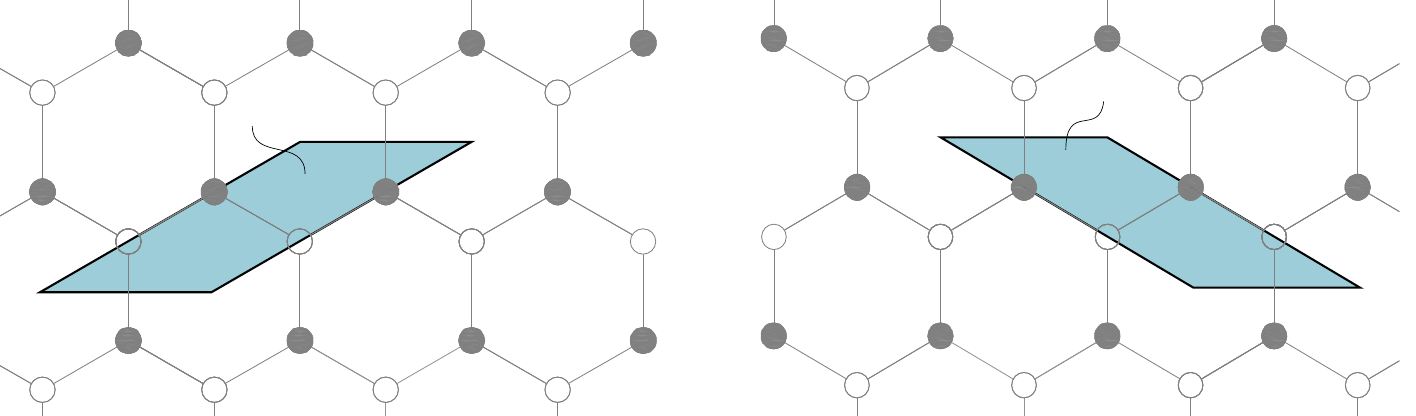
	\caption{ The regions $P^{(1)}_{3\ell}(\xb^\ell)$ and $P^{(1)}_{1\ell}(\xb^\ell)$.}
	\label{fig:P3l}
\end{figure}

We start by studying the behavior of the $\mathcal{Z}$--dihedral energy. 
From \eqref{ThetaZ13} we see that the dihedral angle $\Thz_{\pb_1\pb_{3}}[w_\ell](\xb^\ell) $ is
proportional to $\jump{\nabla\hat w_\ell^{(1)}}_{\pb_1}(\xb^\ell)\cdot \pb_{3}$. Hence, to understand the limit
behavior of the dihedral angles we may study the behavior of particular jumps.
With this in mind, we set
\begin{equation}\label{J13l0}
J^{(1)}_{3\ell}(\xb):=\sum_{\xb^\ell\in L_2}\frac{\jump{\nabla\hat w_\ell^{(1)}}_{\pb_1}(\xb^\ell)\cdot\pb_3}{\sqrt{3}\ell}\chi_{P^{(1)}_{3\ell}(\xb^\ell)}(\xb),
\end{equation}
with $P^{(1)}_{3\ell}(\xb^\ell)$ the parallelograms of height $\frac{3}{2}\ell$ and base $\sqrt{3}\ell$, with sides parallel to $\db_1$ and $\pb_1$ passing through the nodes $\xb^\ell+\ell \pb_1$ and $\xb^\ell+\ell \pb_3$,  cf. Fig.~\ref{fig:P3l}.  

Note that
$\cup_{\xb^\ell\in L_2(\ell)} P^{(1)}_{3\ell}(\xb^\ell)$ is equal to $\Ro^2$, up to a set of measure zero.
By \eqref{ThetaZ13} we may also write
\begin{equation}\label{J13l1}
J^{(1)}_{3\ell}(\xb)=-\sum_{\xb^\ell\in L_2}\frac{1}{2\ell}\Thz_{\pb_1\pb_{3}}[w_\ell](\xb^\ell)\chi_{P^{(1)}_{3\ell}(\xb^\ell)}(\xb).
\end{equation}
Similarly, we set
\begin{equation}\label{J11l0}
J^{(1)}_{1\ell}(\xb):=\sum_{\xb^\ell\in L_2}\frac{\jump{\nabla\hat w_\ell^{(1)}}_{\pb_3}(\xb^\ell)\cdot\pb_1}{\sqrt{3}\ell}\chi_{P^{(1)}_{1\ell}(\xb^\ell)}(\xb),
\end{equation}
with $P^{(1)}_{1\ell}(\xb^\ell)$ the parallelograms with sides parallel to $\db_1$ and $\pb_3$ passing through the nodes $\xb^\ell+\ell \pb_1$ and $\xb^\ell+\ell \pb_3$.

\begin{lem}\label{lemJ1}
	Let $\hat w^{(1)}_\ell$ be as in Lemma~\ref{lem1}, $\gb^{(1)}$ be as in Theorem \ref{gi}, $J^{(1)}_{3\ell}$ and $J^{(1)}_{1\ell}$ be the functions defined in \eqref{J13l0} and \eqref{J11l0}, respectively.
	Then, there exists a subsequence, not relabeled,  such that
	\begin{equation}\label{limJ1}
	J^{(1)}_{3\ell} \rightharpoonup 3 \,\partial_{\db_1}\gb^{(1)}\cdot\pb_3,
	\quad\mbox{and}\quad
	J^{(1)}_{1\ell} \rightharpoonup 3 \,\partial_{\db_1}\gb^{(1)}\cdot\pb_1
	\quad \mbox{in }L^2(\mathbb{R}^2).
	\end{equation}
\end{lem}

\proof 
By integrating \eqref{J13l1} we find that
$$
\int_{\mathbb{R}^2}(J^{(1)}_{3\ell})^2\,d\xb=
\sum_{\xb^\ell\in L_2}\frac{1}{4\ell^2}\Big(\Thz_{\pb_1\pb_{3}}[w_\ell](\xb^\ell)\Big)^2\,\frac{3\sqrt{3}}{2}\ell^2\le C,
$$
where the last inequality follows since, by assumption, $w_\ell$  satisfies the energy bound \eqref{energybound}.
Hence, there exists a  subsequence of $J^{(1)}_{3\ell}$ weakly convergent in $L^2(\Ro^2)$.
We now characterize its limit.

By writing explicitly the jump in \eqref{J13l0} we find
$$
J^{(1)}_{3\ell}(\xb)=\sum_{\xb^\ell\in L_2}\pb_3\cdot\frac{\nabla\hat w_\ell^{(1)}(\xb^\ell+\ell\pb_1)-\nabla\hat w_\ell^{(1)}(\xb^\ell)}{\sqrt{3}\ell}\chi_{P^{(1)}_{3\ell}(\xb^\ell)}(\xb),
$$
which we may rewrite more compactly as
$$
J^{(1)}_{3\ell}(\xb)=\sum_{\xb^\ell\in L_2}\frac{\partial_{\pb_3}\hat w_\ell^{(1)}(\xb^\ell+\ell\pb_1)-\partial_{\pb_3}\hat w_\ell^{(1)}(\xb^\ell)}{\sqrt{3}\ell}\chi_{P^{(1)}_{3\ell}(\xb^\ell)}(\xb).
$$
Let $\varphi\in C^\infty_0(\Ro^2)$. Since on
\begin{equation*}
\bar{P}^{(1)}_{3\ell}(\xb^\ell):=P^{(1)}_{3\ell}(\xb^\ell)\cap cS^{(1)}_\ell,
\end{equation*}
the pointwise derivative $\partial_{\pb_3} \hat w_\ell^{(1)}$ is constant, and since for every $\xb\in \bar{P}^{(1)}_{3\ell}(\xb^\ell)$ we have that $\xb+\ell \db_1\in \bar{P}^{(1)}_{3\ell}(\xb^\ell+\ell \db_1)$,
the identity
$$
\partial_{\pb_3} \hat w_\ell^{(1)}(\xb^\ell+\ell \pb_1)-\partial_{\pb_3}\hat w_\ell^{(1)}(\xb^\ell)=
\partial_{\pb_3} \hat w_\ell^{(1)}(\xb+\ell\db_1)-\partial_{\pb_3}\hat w_\ell^{(1)}(\xb)
$$
holds for every $\xb\in \bar{P}^{(1)}_{3\ell}(\xb^\ell)$.
Then,
\begin{align*}
\int_{\mathbb{R}^2}J^{(1)}_{3\ell}&\varphi\,d\xb =\sum_{\xb^\ell\in L_2}\int_{P^{(1)}_{3\ell}(\xb^\ell)}\frac{\partial_{\pb_3} \hat w_\ell^{(1)}(\xb^\ell+\ell \pb_1)-\partial_{\pb_3}\hat w_\ell^{(1)}(\xb^\ell)}{\sqrt{3}\ell}\varphi(\xb)\,d\xb\nonumber\\
&=\sum_{\xb^\ell\in L_2}\int_{\bar{P}^{(1)}_{3\ell}(\xb^\ell)}\frac{\partial_{\pb_3} \hat w_\ell^{(1)}(\xb+\ell\db_1)-\partial_{\pb_3}\hat w_\ell^{(1)}(\xb)}{\sqrt{3}\ell}\varphi(\xb)\,d\xb\nonumber\\
&\ +\sum_{\xb^\ell\in L_2}\int_{\bar{P}^{(1)}_{3\ell}(\xb^\ell)}\frac{\partial_{\pb_3} \hat w_\ell^{(1)}(\xb+\ell\db_1)-\partial_{\pb_3}\hat w_\ell^{(1)}(\xb)}{\sqrt{3}\ell}\varphi(\xb+\ell/2\pb_1)\,d\xb\nonumber\\
&\ +\sum_{\xb^\ell\in L_2}\int_{\bar{P}^{(1)}_{3\ell}(\xb^\ell)}\frac{\partial_{\pb_3} \hat w_\ell^{(1)}(\xb+\ell\db_1)-\partial_{\pb_3}\hat w_\ell^{(1)}(\xb)}{\sqrt{3}\ell}\varphi(\xb-\ell/2\pb_1)\,d\xb.
\end{align*}
Hence, by a change of variables and a rearrangement of the sums we deduce
\begin{align*}
\int_{\mathbb{R}^2}J^{(1)}_{3\ell}&\varphi\,d\xb =\sum_{\xb^\ell\in L_2}\int_{\bar{P}^{(1)}_{3\ell}(\xb^\ell)}\frac{\varphi(\xb-\ell\db_1)-\varphi(\xb)}{\sqrt{3}\ell}\partial_{\pb_3}\hat w_\ell^{(1)}(\xb)\,d\xb\nonumber\\
&+\sum_{\xb^\ell\in L_2}\int_{\bar{P}^{(1)}_{3\ell}(\xb^\ell)}\frac{\varphi(\xb-\ell\db_1+\ell/2\pb_1)-\varphi(\xb+\ell/2\pb_1)}{\sqrt{3}\ell}\partial_{\pb_3}\hat w_\ell^{(1)}(\xb)\,d\xb\nonumber\\
&+\sum_{\xb^\ell\in L_2}\int_{\bar{P}^{(1)}_{3\ell}(\xb^\ell)}\frac{\varphi(\xb-\ell\db_1-\ell/2\pb_1)-\varphi(\xb-\ell/2\pb_1)}{\sqrt{3}\ell}\partial_{\pb_3}\hat w_\ell^{(1)}(\xb)\,d\xb.
\end{align*}
After observing that $\partial_{\pb_3}\hat w_\ell^{(1)}(\xb)=\gb^{(1)}_\ell(\xb)\cdot\pb_3$ for every $\xb\in cS^{(1)}_\ell$ and recalling \eqref{weakgli}, we pass to the limit to obtain 
$$
\lim_{\ell\to 0}\int_{\mathbb{R}^2}J^{(1)}_{3\ell}\varphi\,d\xb = 3\int_{\mathbb{R}^2}-\partial_1\varphi \,\,\gb^{(1)}\cdot\pb_3\,d\xb, \quad \forall \varphi\in C^\infty_0(\Ro^2),
$$
that is,
$$
J^{(1)}_{3\ell} \rightharpoonup 3 \,\partial_{\db_1}\gb^{(1)}\cdot\pb_3 \quad \mbox{in $L^2(\Ro^2)$}.
$$
The statement about $J^{(1)}_{1\ell}$ is proved similarly. 
\QED

\begin{rem} \label{rem1}
	To  contain  the notation, in Lemma \ref{lemJ1} we stated the result just for the jumps of $\hat w^{(1)}_\ell$.
	But similarly, we may define the functions $J^{(2)}_{2\ell}$ and $J^{(2)}_{3\ell} $ for the piecewise affine interpolant $\hat w_\ell^{(2)}$ along the nodes of $cS_\ell^{(2)}$, and the functions $J^{(3)}_{1\ell}$ and $J^{(3)}_{2\ell}$ for the interpolant $\hat w_\ell^{(3)}$ along $cS_\ell^{(3)}$, and find
	\begin{equation}\label{limJ23}
	\begin{array}{c}
	J^{(2)}_{3\ell} \rightharpoonup 3 \,\partial_{\db_2}\gb^{(2)}\cdot\pb_3 \quad \mbox{and} \quad J^{(2)}_{2\ell} \rightharpoonup 3 \,\partial_{\db_2}\gb^{(2)}\cdot\pb_2, \\[4pt]
	J^{(3)}_{1\ell} \rightharpoonup 3 \,\partial_{\db_3}\gb^{(3)}\cdot\pb_1 \quad \mbox{and} \quad J^{(3)}_{2\ell} \rightharpoonup 3 \,\partial_{\db_3}\gb^{(2)}\cdot\pb_2,
	\end{array}
	\end{equation}
	in $L^2(\Ro^2)$.
\end{rem}

The next lemma deals with the regularity of $w$ and $\gamma$.
\begin{lem}\label{regularity}
	Let  $w$ be as in Theorem \ref{thm-wstrong} and $\gamma$ as in Theorem \ref{gi}.
	Then $w\in H^2(\Ro^2)$, $\gamma\in H^1(\Ro^2)$, and both functions are 
	equal to zero almost everywhere outside of $\Omega$.
\end{lem}
\proof
We already know that $w\in H^1(\Ro^2)$, $\gamma\in L^2(\Ro^2)$, and that both functions are 
equal to zero almost everywhere outside of $\Omega$, {\it cf.} Theorem \ref{thm-wstrong}, Lemma \ref{lemwhw0}, and 
Theorem \ref{gi}. By Lemma \ref{lemJ1},
$$
\partial_{\db_1}\gb^{(1)}\cdot\pb_3\in L^2(\mathbb{R}^2),\quad\mbox{and}\quad
\partial_{\db_1}\gb^{(1)}\cdot\pb_1
\in L^2(\mathbb{R}^2),
$$
and hence $\partial_{\db_1}\gb^{(1)}\in L^2(\mathbb{R}^2;\Ro^2)$. Similarly by Remark \ref{rem1}, we deduce that
$$
\partial_{\db_i}\gb^{(i)}\in L^2(\mathbb{R}^2;\Ro^2), \qquad i=1,2,3.
$$
By scalar multiplication by $\db_i$ and $\db_i^\perp$, this and \eqref{girep} imply that
$$
\partial_{\db_i\db_i}w\in L^2(\Ro^2), \quad \partial_{\db_i}\gamma\in L^2(\Ro^2)
$$
for $i=1,2,3.$ This implies $w\in H^2(\Ro^2)$ and $\gamma\in H^1(\Ro^2)$.
\QED

\noindent
{\sc Proof of Theorem  \ref{thm-wstrong_0}.}
The proof follows by putting together Theorem~\ref{thm-wstrong} and Lemma~\ref{regularity}.
\QED

We now prove a lower bound for the $\liminf$ of the $\mathcal{Z}$-- dihedral energy. 

\begin{lem}\label{thm:liminfZ}
	Let $w_\ell\in \mathcal{A_\ell}$ satisfy the energy bound \eqref{energybound}, let $w\in H^2(\Ro^2)$ and $\gamma\in H^1(\Ro^2)$ be as in Theorem \ref{regularity}. Then, 
	$$
	\liminf_{\ell\to 0} \mathcal{U}^{\mathcal{Z}}_\ell(w_\ell)\ge \mathcal{U}^{\mathcal{Z}}_0(w, \gamma),
	$$
	where 
	\begin{align}
	\mathcal{U}^{\mathcal{Z}}_0(w, \gamma):=
	\frac{4\sqrt{3}}{9}\,k^{\mathcal{Z}}\int_{\mathbb{R}^2}&(\partial_{\db_1\pb_3}w-\frac{3}{2}\partial_{\db_1}\gamma)^2+(\partial_{\db_1\pb_1}w-\frac{3}{2}\partial_{\db_1}\gamma)^2\nonumber\\
	&+(\partial_{\db_2\pb_3}w-\frac{3}{2}\partial_{\db_2}\gamma)^2+(\partial_{\db_2\pb_2}w-\frac{3}{2}\partial_{\db_2}\gamma)^2\label{Uz0}\\
	&+(\partial_{\db_3\pb_1}w-\frac{3}{2}\partial_{\db_3}\gamma)^2+(\partial_{\db_3\pb_2}w-\frac{3}{2}\partial_{\db_3}\gamma)^2\,d\xb.\nonumber
	\end{align}
\end{lem}

\proof 
Consider the first term of \eqref{Z-split} and use \eqref{J13l1} to find that
\begin{align*}
\sum_{\xb^\ell\in L_2(\ell)} 
\Big(\Thz_{\pb_1\pb_{3}}[w_\ell](\xb^\ell)\Big)^2=\frac{8\sqrt{3}}{9}\int_{\mathbb{R}^2}(J^{(1)}_{3\ell})^2\,d\xb.
\end{align*}
Hence, by \eqref{limJ1},
$$
\liminf_{\ell\to 0} \frac 12 k^{\mathcal{Z}}\sum_{\xb^\ell\in L_2(\ell)} 
\Big(\Thz_{\pb_1\pb_{3}}[w_\ell](\xb^\ell)\Big)^2
\ge 
\frac{4\sqrt{3}}{9}\, k^{\mathcal{Z}} \int_{\mathbb{R}^2}  (3 \,\partial_{\db_1}\gb^{(1)}\cdot\pb_3)^2\,d\xb.
$$
Since all the other terms of  \eqref{Z-split} can be treated similarly, we find that
\begin{align*}
\liminf_{\ell\to 0} \mathcal{U}^{\mathcal{Z}}_\ell(w_\ell)
&\ge \nonumber \\
&\hspace{-10mm}4\sqrt{3}\,k^{\mathcal{Z}} \int_{\mathbb{R}^2}(\partial_{\db_1}\gb^{(1)}\cdot\pb_3)^2+(\partial_{\db_1}\gb^{(2)}\cdot\pb_1)^2+(\partial_{\db_2}\gb^{(2)}\cdot\pb_3)^2\nonumber \\
&\hspace{5mm}+(\partial_{\db_2}\gb^{(2)}\cdot\pb_2)^2+(\partial_{\db_3}\gb^{(3)}\cdot\pb_1)^2+(\partial_{\db_3}\gb^{(3)}\cdot\pb_2)^2\,d\xb.
\end{align*}
Thanks to \eqref{girep}, the integral on the right hand side is equal to $\mathcal{U}^{\mathcal{Z}}_0(w, \gamma)$.
\QED

\begin{rem} \label{rem2}
	We notice that, by expressing the derivatives that appear in \eqref{Uz0} in terms of partial derivatives with respect to Cartesian coordinates, it is possible to show that $\mathcal{U}^{\mathcal{Z}}_0(w, \gamma)$ coincides with the functional $\mathcal{U}^{\mathcal{Z}(b)}_0(w, \gamma)$ introduced in \eqref{Uz00}, see \cite{Davini_2017} for the details.
\end{rem}

We now prove a lower bound for the $\liminf$ of the $\mathcal{C}$-- dihedral energy. 

\begin{lem}\label{thm:liminfC}
	Let $w_\ell\in \mathcal{A_\ell}$ satisfy the energy bound \eqref{energybound}, let $w\in H^2(\Ro^2)$ be as in Theorem \ref{regularity}. Then, 
	$$
	\liminf_{\ell\to 0} \mathcal{U}^{\mathcal{C}}_\ell(w_\ell)\ge \mathcal{U}^{\mathcal{C}}_0(w),
	$$
	where 
	\begin{align}
	\mathcal{U}^{\mathcal{C}}_0(w):=\frac{8\sqrt{3}}{9}k^{\mathcal{C}}\int_{\Ro^2}\sum_{i=1}^3(
	\partial_{\pb_i\pb_i^\perp}w)^2\,d\xb.\label{Uc0}
	\end{align}
\end{lem}

\proof 
For $k^{\mathcal{C}}=0$ the statement of the lemma trivially holds. Hence, we suppose $k^{\mathcal{C}}>0$.

\begin{figure}[h]
	\centering
	\def\svgwidth{.35\textwidth}
	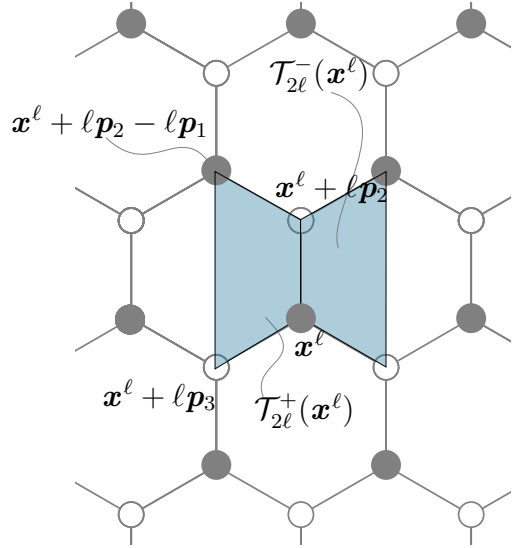
	\caption{ The trapezoids $\mathcal{T}_{2\ell}^+(\xb^\ell)$ and  $\mathcal{T}_{2\ell}^-(\xb^\ell)$.}
	\label{fig:trapezoid}
\end{figure}

For $\xb^\ell\in L_2(\ell)$,
denote by $\mathcal{T}_{2\ell}^+(\xb^\ell)$  the trapezoid with one base the bond edge starting at $\xb^\ell$ and ending at $\xb^\ell+\ell\pb_2$ and the other base the segment joining the points $\xb^\ell+\ell\pb_3$  and $\xb^\ell+\ell\pb_2-\ell\pb_1$. Let $\mathcal{T}_{2\ell}^-(\xb^\ell)$ be the trapezoid obtained by reflecting $\mathcal{T}_{2\ell}^+(\xb^\ell)$ with respect to the bond edge starting at $\xb^\ell$ and parallel to $\pb_2$, see Fig. \ref{fig:trapezoid}.

Denote by  $A_\ell:=\frac{3\sqrt{3}}{4}\ell^2$ the area of the trapezoids, and let 
\begin{equation}\label{pcl2}
\pcl2(\xb):=\sum_{\xb^\ell\in L_2}\frac{1}{\sqrt{A_\ell}}\Big(\Thc_{\pb_2^+}[w_\ell](\xb^\ell)\chi_{\mathcal{T}_{2\ell}^+(\xb^\ell)}(\xb)-\Thc_{\pb_2^-}[w_\ell](\xb^\ell)\chi_{\mathcal{T}_{2\ell}^-(\xb^\ell)}(\xb)\Big).
\end{equation}
It follows that
\begin{equation}\label{use0}
\int_{\Ro^2}|\pcl2|^2\,d\xb=\sum_{\xb^\ell\in L_2(\ell)}\Big(\Thc_{\pb_2^+}[w_\ell](\xb^\ell)\Big)^2+\Big(\Thc_{\pb_2^-}[w_\ell](\xb^\ell)\Big)^2\le C,
\end{equation}
where the last inequality follows since $w_\ell$ satisfies the energy bound \eqref{energybound} and because  $k^{\mathcal{C}}>0$, see \eqref{enC}.
Then, up to a subsequence, $\pcl2$ weakly converges to some $\pc2\in L^2(\Ro^2)$.

Hereafter we characterize $\pc2$.
From \eqref{C-wedge} evaluated for $i=2$, we find
\begin{align}
\Thc_{\pb_2^+}[w_\ell](\xb^\ell)&=\frac{2\sqrt{3}}{3\ell}\Big[w_\ell(\xb^\ell)-w_\ell(\xb^\ell+\ell \pb_{3})-w_\ell(\xb^\ell+\ell \pb_{2})\nonumber\\
&\hspace{1cm}-\Big(w_\ell(\xb^\ell+\ell \pb_{2})-w_\ell(\xb^\ell+\ell\pb_2-\ell \pb_{1})-w_\ell(\xb^\ell)\Big)\Big],\nonumber\\
&=\frac{2\sqrt{3}}{3\ell}\Big[\hat w^{(2)}_\ell(\xb^\ell)-\hat w^{(2)}_\ell(\xb^\ell+\ell \pb_{3})-\hat w^{(2)}_\ell(\xb^\ell+\ell \pb_{2})\nonumber\\
&\hspace{1cm}-\Big(\hat w^{(3)}_\ell(\xb^\ell+\ell \pb_{2})-\hat w^{(3)}_\ell(\xb^\ell+\ell\pb_2-\ell \pb_{1})-\hat w^{(3)}_\ell(\xb^\ell)\Big)\Big],\nonumber\\
&=\Big(\nabla\hat w^{(3)}_\ell(\xb^\ell+\ell \pb_{2})-\nabla \hat w^{(2)}_\ell(\xb^\ell)\Big)\cdot (+\pb_2^\perp),\label{cp+}
\end{align}
where the last identity follows by an easy calculation.	
Similarly, one finds
\begin{equation}\label{cp-}
\Thc_{\pb_2^-}[w_\ell](\xb^\ell)=\Big(\nabla\hat w^{(2)}_\ell(\xb^\ell+\ell \pb_{2})-\nabla \hat w^{(3)}_\ell(\xb^\ell)\Big)\cdot (-\pb_2^\perp).
\end{equation}

Let
$$
\mathcal{A}_{2\ell}(\xb^\ell):=\mathcal{T}_{2\ell}^+(\xb^\ell)\cup\mathcal{T}_{2\ell}^-(\xb^\ell).
$$
Then, with \eqref{pcl2}, \eqref{cp+}, and \eqref{cp-}, we may compute
\begin{align*}
\int_{\mathcal{A}_{2\ell}(\xb^\ell)}\pcl2\,d\xb&=
\sqrt{A_\ell}\Big(\Thc_{\pb_2^+}[w_\ell](\xb^\ell)-\Thc_{\pb_2^-}[w_\ell](\xb^\ell)\Big)\\
&=
\sqrt{A_\ell}\Big(
\nabla\hat w^{(3)}_\ell(\xb^\ell+\ell \pb_{2})-\nabla \hat w^{(2)}_\ell(\xb^\ell)\\
&\hspace{3cm}+\nabla\hat w^{(2)}_\ell(\xb^\ell+\ell \pb_{2})-\nabla \hat w^{(3)}_\ell(\xb^\ell)
\Big)\cdot\pb_2^\perp\\
&=
\sqrt{A_\ell}\Big(\jump{\nabla\hat w^{(3)}_\ell}_{\pb_2}
(\xb^\ell)+\jump{\nabla \hat w^{(2)}_\ell}_{\pb_2}(\xb^\ell)\Big)\cdot\pb_2^\perp.
\end{align*}
Since 
$$
\pb_2^\perp=\frac{\sqrt{3}}3\pb_2+\frac{2\sqrt{3}}3\pb_3=-\frac{\sqrt{3}}3\pb_2-\frac{2\sqrt{3}}3\pb_1,
$$
and since, by continuity, $\jump{\nabla\hat w^{(3)}_\ell}_{\pb_2}
(\xb^\ell)\cdot\pb_2=\jump{\nabla \hat w^{(2)}_\ell}_{\pb_2}(\xb^\ell)\cdot\pb_2=0$, we have that
\begin{align*}
\int_{\mathcal{A}_{2\ell}(\xb^\ell)}\pcl2\,d\xb&=
-\frac{2\sqrt{3}}3\sqrt{A_\ell}\Big(\jump{\nabla\hat w^{(3)}_\ell}_{\pb_2}
(\xb^\ell)\cdot\pb_1-\jump{\nabla \hat w^{(2)}_\ell}_{\pb_2}(\xb^\ell)\cdot\pb_3\Big).
\end{align*}
Let $P^{(3)}_{1\ell}(\xb^\ell)$ and $P^{(2)}_{3\ell}(\xb^\ell)$ be the rhomboidal regions used in the definition of the functions $J^{(3)}_{1\ell}$
and $J^{(2)}_{3\ell}$, respectively.  Since the area of these regions are equal to $3\sqrt{3}\ell^2/2$ we may write
\begin{align}
\int_{\mathcal{A}_{2\ell}(\xb^\ell)}\pcl2\,d\xb&=
-\frac{2\sqrt{3\sqrt{3}}}{9\ell}\Big(\int_{P^{(3)}_{1\ell}(\xb^\ell)}\jump{\nabla\hat w^{(3)}_\ell}_{\pb_2}
(\xb^\ell)\cdot\pb_1\,d\xb\nonumber\\
&\hspace{3cm}-\int_{P^{(2)}_{3\ell}(\xb^\ell)}\jump{\nabla \hat w^{(2)}_\ell}_{\pb_2}(\xb^\ell)\cdot\pb_3\,d\xb\Big)\nonumber\\
&=
-\frac{2}{\sqrt{3\sqrt{3}}}\Big(\int_{P^{(3)}_{1\ell}(\xb^\ell)}J^{(3)}_{1\ell}\,d\xb
-\int_{P^{(2)}_{3\ell}(\xb^\ell)}J^{(2)}_{3\ell}\,d\xb\Big).\label{use1}
\end{align}

Let $B\subset \Ro^2$ be an open set, and let 
\begin{align*}
B_\ell&:=\{\cup_{\xb^\ell\in L_2(\ell)} \mathcal{A}_{2\ell}(\xb^\ell): \mathcal{A}_{2\ell}(\xb^\ell)\subset B\},\\
B^{(2)}_\ell&:=\{\cup_{\xb^\ell\in L_2(\ell)} P^{(2)}_{3\ell}(\xb^\ell): \mathcal{A}_{2\ell}(\xb^\ell)\subset B\},\\
B^{(3)}_\ell&:=\{\cup_{\xb^\ell\in L_2(\ell)} P^{(3)}_{1\ell}(\xb^\ell): \mathcal{A}_{2\ell}(\xb^\ell)\subset B\}.
\end{align*}
Then, by \eqref{use1} we have
\begin{align}
\int_B \pc2\,d\xb&=\lim_{\ell\to 0}\int_B \pcl2\,d\xb=\lim_{\ell\to 0} \int_{B_\ell}\pcl2\,d\xb\nonumber\\
&=-\frac{2}{\sqrt{3\sqrt{3}}}\lim_{\ell\to 0}\Big(\int_{B^{(3)}_{\ell}}J^{(3)}_{1\ell}\,d\xb
-\int_{B^{(2)}_{\ell}}J^{(2)}_{3\ell}\,d\xb\Big)\nonumber\\
&=-\frac{2}{\sqrt{3\sqrt{3}}}\int_{B}
3 \,\partial_{\db_3}\gb^{(3)}\cdot\pb_1-3 \,\partial_{\db_2}\gb^{(2)}\cdot\pb_3
\,d\xb.\label{use2}
\end{align}

By taking the characterization of the $\gb^{(i)}$ into account, see \eqref{girep}, it follows that
\begin{equation*}\label{pcl2wg}
3(\partial_{\db_2}\gb^{(2)}\cdot\pb_3- \partial_{\db_3} \gb^{(3)}\cdot\pb_1)= (\partial_{\db_2\pb_3}w - \partial_{\db_3\pb_1}w - \frac{3}{2}(\partial_{\db_2}\gamma - \partial_{\db_3}\gamma)).
\end{equation*}
But $\gamma=0$ by Theorem \ref{gamma}. Furthermore, as it is easily seen, the following relations hold:
$$
\partial_{\db_2\pb_3}w - \partial_{\db_3\pb_1}w=2\partial_{\pb_2\pb_2^\perp}w.
$$
Thus, \eqref{use2} rewrites as
\begin{equation*}
\int_B \pc2\,d\xb=\frac{4}{\sqrt{3\sqrt{3}}}\int_{B}
\partial_{\pb_2\pb_2^\perp}w 
\,d\xb,
\end{equation*}
and since this identity holds for every open set $B$, we deduce that
$$
\pc2=\frac{4}{\sqrt{3\sqrt{3}}}\,
\partial_{\pb_2\pb_2^\perp}w .
$$
Finally, from \eqref{use0}
\begin{align*}
\liminf_{\ell\to 0}\sum_{\xb^\ell\in L_2(\ell)}\Big(&\Thc_{\pb_2^+}[w_\ell](\xb^\ell)\Big)^2+\Big(\Thc_{\pb_2^-}[w_\ell](\xb^\ell)\Big)^2=\liminf_{\ell\to 0}\int_{\Ro^2}|\pcl2|^2\,d\xb\\
&\ge\int_{\Ro^2}|\pc2|^2\,d\xb=\frac{16\sqrt{3}}{9}\int_{\Ro^2}(
\partial_{\pb_2\pb_2^\perp}w)^2\,d\xb.
\end{align*}
Similar inequalities can be proved also for $\pb_1$  and $\pb_3$; hence, from \eqref{enC} we deduce the statement of the Lemma.
\QED

Next, we consider the self-energy.

\begin{lem}\label{thm:liminfs}
	Let $w_\ell\in \mathcal{A_\ell}$ satisfy the energy bound \eqref{energybound}, let $w\in H^2(\Ro^2)$  be as in Theorem \ref{regularity}. Then, 
	$$
	\liminf_{\ell\to 0} \mathcal{U}^{s}_\ell(w_\ell)\ge \mathcal{U}^{s}_0(w),
	$$
	where 
	\begin{align}
	\mathcal{U}^{s}_0(w):=\frac{-\tau_0}{18}\int_{\Ro^2}\Big( \sum_{i=1}^3\partial_{\pb_i\pb_i}w\Big)^2\,d\xb.
	\label{Us0}
	\end{align}
\end{lem}

\proof
Since the inequality trivially holds for $\tau_0=0$, we may assume $\tau_0<0$. Set
$$
\psl(\xb):=\frac{2}{\sqrt{3\sqrt{3}}\ell}\Big(\sum_{\xb^\ell\in L_1(\ell)} \Ths_1[w_\ell](\xb^\ell)\chi_{T^\ell(\xb^\ell)}(\xb)+\sum_{\xb^\ell\in L_2(\ell)}\Ths_2[w_\ell](\xb^\ell)\chi_{T^\ell(\xb^\ell)}(\xb)\Big).$$
Then, since the area of $T^\ell(\xb^\ell)$ is equal to $3\sqrt{3}\ell^2/4$, with \eqref{ens},
\begin{align}
\int_{\Ro^2}|\psl|^2\,d\xb&=\sum_{\xb^\ell\in L_1(\ell)} 
\, \Big(\Ths_1[w_\ell](\xb^\ell)\Big)^2+\sum_{\xb^\ell\in L_2(\ell)} 
\, \Big(\Ths_2[w_\ell](\xb^\ell)\Big)^2\label{suse0}\\
&=\frac{-2\mathcal{U}^{s}_\ell(w_\ell)}{\tau_0}\le C,\nonumber
\end{align}
where the inequality follows because $w_\ell$ satisfies the energy bound \eqref{energybound}. Hence, up to a subsequence,
$$
\psl\rightharpoonup \ps\quad\mbox{in }L^2(\Ro^2),
$$
for some $\ps\in L^2(\Ro^2)$.

From \eqref{s_angles} we compute the strain measures
\begin{align}
\Ths_1[w_\ell](\xb^\ell)&=\frac{\sqrt{3\sqrt{3}}}{3\ell}\Big(\hat w^{(3)}_\ell(\xb^\ell-\ell\pb_1)-\hat w^{(3)}_\ell(\xb^\ell)\nonumber\\
&\hspace{3cm}+\hat w^{(2)}_\ell(\xb^\ell-\ell\pb_2)-\hat w^{(2)}_\ell(\xb^\ell)\nonumber\\
&\hspace{3cm}+\hat w^{(1)}_\ell(\xb^\ell-\ell\pb_3)-\hat w^{(1)}_\ell(\xb^\ell)\Big),\nonumber\\
&=-\frac{\sqrt{3\sqrt{3}}}{3}\Big(\partial_{\pb_1}\hat w^{(3)}_\ell(\xb^\ell)+\partial_{\pb_2}\hat w^{(2)}_\ell(\xb^\ell)+\partial_{\pb_3}\hat w^{(1)}_\ell((\xb^\ell)\Big)\nonumber
.\label{suse1}
\end{align}
and
\begin{equation*}
\Ths_2[w_\ell](\xb^\ell)=\frac{\sqrt{3\sqrt{3}}}{3}\Big(\partial_{\pb_1}\hat w^{(3)}_\ell(\xb^\ell)+\partial_{\pb_2}\hat w^{(2)}_\ell(\xb^\ell)+\partial_{\pb_3}\hat w^{(1)}_\ell((\xb^\ell)\Big).
\end{equation*}
Therefore the function $\psl$ takes the form
\begin{align*}
\psl(\xb)&=\frac{2}{3\ell}\Big(\sum_{\xb^\ell\in L_2(\ell)}\Big(\partial_{\pb_1}\hat w^{(3)}_\ell(\xb^\ell)+\partial_{\pb_2}\hat w^{(2)}_\ell(\xb^\ell)+\partial_{\pb_3}\hat w^{(1)}_\ell((\xb^\ell)\Big)\chi_{T^\ell(\xb^\ell)}(\xb)\nonumber\\
&\hspace{1cm}
-\sum_{\xb^\ell\in L_1(\ell)}\Big(\partial_{\pb_1}\hat w^{(3)}_\ell(\xb^\ell)+\partial_{\pb_2}\hat w^{(2)}_\ell(\xb^\ell)+\partial_{\pb_3}\hat w^{(1)}_\ell((\xb^\ell)\Big)\chi_{T^\ell(\xb^\ell)}(\xb)
\Big)\\
&=\frac{2}{3}\big(\ps_{1\ell}(\xb)+\ps_{2\ell}(\xb)+\ps_{3\ell}(\xb)\big),
\end{align*}
where we have set
\begin{align*}
\ps_{1\ell}(\xb)&:=\frac 1\ell \Big(\sum_{\xb^\ell\in L_2(\ell)}\partial_{\pb_1}\hat w^{(3)}_\ell(\xb^\ell)\chi_{T^\ell(\xb^\ell)}(\xb)
-\sum_{\xb^\ell\in L_1(\ell)}\partial_{\pb_1}\hat w^{(3)}_\ell(\xb^\ell)\chi_{T^\ell(\xb^\ell)}(\xb)\Big),\\
\ps_{2\ell}(\xb)&:=\frac 1\ell \Big(\sum_{\xb^\ell\in L_2(\ell)}\partial_{\pb_2}\hat w^{(2)}_\ell(\xb^\ell)\chi_{T^\ell(\xb^\ell)}(\xb)
-\sum_{\xb^\ell\in L_1(\ell)}\partial_{\pb_2}\hat w^{(2)}_\ell(\xb^\ell)\chi_{T^\ell(\xb^\ell)}(\xb)\Big),\\
\ps_{3\ell}(\xb)&:=\frac 1\ell \Big(\sum_{\xb^\ell\in L_2(\ell)}\partial_{\pb_3}\hat w^{(1)}_\ell(\xb^\ell)\chi_{T^\ell(\xb^\ell)}(\xb)
-\sum_{\xb^\ell\in L_1(\ell)}\partial_{\pb_3}\hat w^{(1)}_\ell(\xb^\ell)\chi_{T^\ell(\xb^\ell)}(\xb)\Big).
\end{align*}
Let $\varphi\in C_0^\infty(\Ro^2)$. Then
\begin{align}
\int_{\Ro^2}\psl\varphi\,d\xb=\frac{2}{3}\sum_{i=1}^3\int_{\Ro^2}\ps_{i\ell}\varphi\,d\xb.\label{suse3}
\end{align}
We focus on the case $i=3$, the other cases are treated similarly. We have
\begin{align}
\int_{\Ro^2}\ps_{3\ell}\varphi\,d\xb & = \frac{1}{\ell}\Big(\sum_{\xb^\ell\in L_2(\ell)}\partial_{\pb_3}\hat w^{(1)}_\ell(\xb^\ell)\int_{T^\ell(\xb^\ell)}\varphi\,d\xb\nonumber\\
&\hspace{2cm}-\sum_{\xb^\ell\in L_1(\ell)}\partial_{\pb_3}\hat w^{(1)}_\ell(\xb^\ell)\int_{T^\ell(\xb^\ell)}\varphi\,d\xb\Big)\nonumber\\
&=
\frac{1}{\ell}\Big(\sum_{\xb^\ell\in L_2(\ell)}\partial_{\pb_3}\hat w^{(1)}_\ell(\xb^\ell)\int_{T^\ell(\xb^\ell)}\varphi\,d\xb
\nonumber\\
&\hspace{2cm}-\partial_{\pb_3}\hat w^{(1)}_\ell(\xb^\ell+\ell\pb_3)\int_{T(\xb^\ell+\ell\pb_3)}\varphi\,d\xb\Big)
\nonumber\\
&=
\frac{1}{\ell}\sum_{\xb^\ell\in L_2(\ell)}\partial_{\pb_3}\hat w^{(1)}_\ell(\xb^\ell)\Big(\int_{T^\ell(\xb^\ell)}\varphi\,d\xb
-\int_{T(\xb^\ell+\ell\pb_3)}\varphi\,d\xb\Big),
\nonumber
\end{align}
where we have used that $\partial_{\pb_3}\hat w^{(1)}_\ell(\xb^\ell)=\partial_{\pb_3}\hat w^{(1)}_\ell(\xb^\ell+\ell\pb_3)$.

By using Taylor's expansion theorem we get   
$$
\varphi(\xb)= \varphi(\xb^\ell)+\nabla\varphi(\xb^\ell)\cdot(\xb-\xb^\ell)+O(|\xb-\xb^\ell|^2),
$$
whence
\begin{align*}
\frac{1}{\ell}\Big(\int_{T^\ell(\xb^\ell)}\varphi\,d\xb
-&\int_{T(\xb^\ell+\ell\pb_3)}\varphi\,d\xb\Big)=\frac{3\sqrt{3}\ell^2}{4}\Big(-\nabla\varphi(\xb^\ell)\cdot\pb_3+O(\ell)\Big)\\[3pt]
&=\frac{3}{2}\int_{T^{(1)}(\xb^\ell)\cup T^{(1)}(\xb^\ell+\ell\pb_3)}-\partial_{\pb_3}\varphi(\xb^\ell)\,d\xb+O(\ell^3)\\
&=-\frac{3}{2}\int_{T^{(1)}(\xb^\ell)\cup T^{(1)}(\xb^\ell+\ell\pb_3)}\partial_{\pb_3}\varphi(\xb)\,d\xb+O(\ell^3).
\end{align*}
Taking into account that $\partial_{\pb_3}\hat w^{(1)}_\ell$ is constant on $T^{(1)}(\xb^\ell)\cup T^{(1)}(\xb^\ell+\ell\pb_3)$, we have
\begin{align*}
\int_{\Ro^2}\ps_{3\ell}\varphi\,d\xb &=
-\frac{3}{2}\sum_{\xb^\ell\in L_2(\ell)}\int_{T^{(1)}(\xb^\ell)\cup T^{(1)}(\xb^\ell+\ell\pb_3)}\partial_{\pb_3}\hat w^{(1)}_\ell(\xb)\partial_{\pb_3}\varphi(\xb)\,d\xb+O(\ell)
\nonumber\\
&=-\frac{3}{2}\int_{cS^{(1)}_\ell}\partial_{\pb_3}\hat w^{(1)}_\ell(\xb)\partial_{\pb_3}\varphi(\xb)\,d\xb+O(\ell)
\nonumber\\
&=-\frac{3}{2}\int_{\Ro^2}\pb_3\cdot\gb^{(1)}_\ell(\xb)\partial_{\pb_3}\varphi(\xb)\,d\xb+O(\ell),
\end{align*}
that leads to
$$
\lim_{\ell\to 0} \int_{\Ro^2}\ps_{3\ell}\varphi\,d\xb=-\frac{3}{2}\int_{\Ro^2}\pb_3\cdot\gb^{(1)}\partial_{\pb_3}\varphi\,d\xb
=\frac{3}{2}\int_{\Ro^2}\pb_3\cdot\partial_{\pb_3}\gb^{(1)}\varphi\,d\xb.
$$
Similarly, 
\begin{align*}
&\lim_{\ell\to 0} \int_{\Ro^2}\ps_{1\ell}\varphi\,d\xb=\frac{3}{2}\int_{\Ro^2}\pb_1\cdot\partial_{\pb_1}\gb^{(3)}\varphi\,d\xb,\\
&\lim_{\ell\to 0} \int_{\Ro^2}\ps_{2\ell}\varphi\,d\xb=\frac{3}{2}\int_{\Ro^2}\pb_2\cdot\partial_{\pb_2}\gb^{(2)}\varphi\,d\xb,
\end{align*}
and, from \eqref{suse3}, it follows that
\begin{align*}
\ps&=\pb_1\cdot\partial_{\pb_1}\gb^{(3)}+\pb_2\cdot\partial_{\pb_2}\gb^{(2)}+\pb_1\cdot\partial_{\pb_1}\gb^{(3)}\\
&=\sum_{i=1}^3\frac 13 \partial_{\pb_i\pb_i}w-\frac 12\partial_{\pb_i}\gamma=\frac 13 \sum_{i=1}^3\partial_{\pb_i\pb_i}w
\end{align*}
where we used \eqref{girep} and the fact that $\gamma=0$ since we have assumed $\tau_0<0$, cf. Theorem  \ref {gamma}. Finally, from \eqref{suse0} we find
\begin{align*}
\liminf_{\ell\to 0} \mathcal{U}^{s}_\ell(w_\ell)&=\liminf_{\ell\to 0} \frac{-\tau_0}{2}\int_{\Ro^2}|\psl|^2\,d\xb\\
&\ge\frac{-\tau_0}{2}\int_{\Ro^2}|\ps|^2\,d\xb=\frac{-\tau_0}{18}\int_{\Ro^2}\Big( \sum_{i=1}^3\partial_{\pb_i\pb_i}w\Big)^2\,d\xb.
\end{align*}
\QED

\section{Proofs of Theorems \ref{Gamma} and \ref{Gamma2}} \label{proof23}

In this final section we prove that the lower bounds obtained in Section~\ref{sec:Zdihedr} 
provide in fact the $\Gamma$-limit of the energy functional in the two cases envisaged in 
Theorems~\ref{Gamma} and \ref{Gamma2}.
To accomplish the task we need to show that the lower bounds can be achieved;
in the case of smooth target functions, this is done in the next Lemma.

\begin{lem}\label{rec0}
	Let $w\in C^\infty(\Ro^2)$ and $\gamma\in C^\infty(\Ro^2)$ be two functions with support in $\Omega$.
	Then, 
	\begin{enumerate}
		\item
		if $k^{\mathcal{C}}=\tau_0=0$,
		there exists a $w_\ell\in \mathcal{A}_\ell$ such that
		$
		w_\ell \to w\mbox{ in }L^2(\Ro^2),
		$
		and
		$$
		\lim_{\ell\to 0} \mathcal{U}_\ell(w_\ell)=\mathcal{U}_0^{\mathcal{Z}}(w, \gamma),
		$$
		with $\mathcal{U}_0^{\mathcal{Z}}(w,\gamma)$ defined in \eqref{Uz0};
		\item
		if either $k^{\mathcal{C}}\ne0$ or $\tau_0\ne0$,
		there exists a $w_\ell\in \mathcal{A}_\ell$ such that
		$
		w_\ell \to w\mbox{ in }L^2(\Ro^2),
		$
		and
		$$
		\lim_{\ell\to 0} \mathcal{U}_\ell(w_\ell)=\mathcal{U}_0(w),
		$$
		with
		\begin{equation}\label{U00}
		\mathcal{U}_0(w):=\mathcal{U}^{\mathcal{Z}}_0(w, 0)+\mathcal{U}^{\mathcal{C}}_0(w)+\mathcal{U}^{s}_0(w),
		\end{equation} 
		cf. \eqref{Uz0}, \eqref{Uc0}, and \eqref{Us0}.
	\end{enumerate}
\end{lem}

\proof
We start by proving $\it 1.$
Let 
$$
w_\ell(\xb):=\sum_{\xb^\ell\in L_1(\ell)}w(\xb^\ell)\ \chi_{T^\ell(\xb^\ell)}(\xb)+
\sum_{\xb^\ell\in L_2(\ell)}\Big(w(\xb^\ell)+\frac 32 \ell\gamma(\xb^\ell)\Big)\ \chi_{T^\ell(\xb^\ell)}(\xb).
$$
Then, $w_\ell \to w$ in $L^2(\Ro^2)$ and for $\ell$ small enough $w_\ell\in  \mathcal{A}_\ell$ . 

\begin{figure}[h]
	\centering
	\def\svgwidth{.6\textwidth}
	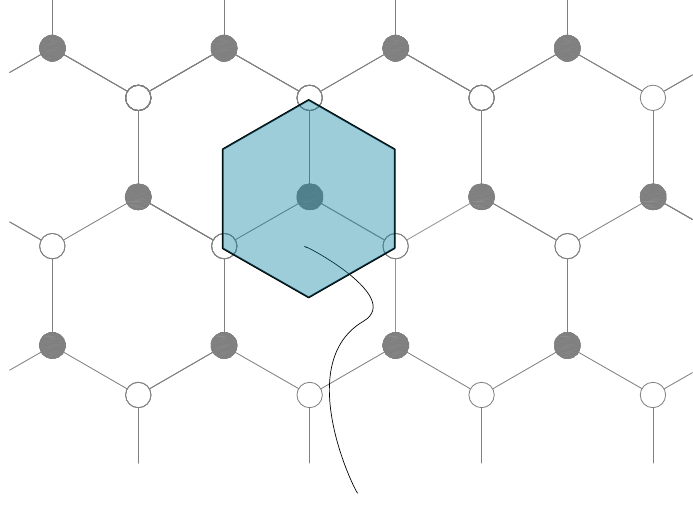
	\caption{The hexagon $E^\ell(\xb^\ell)$.}
	\label{fig:hexagon}
\end{figure}

Recalling \eqref{Z-wedge}, for $\xb^\ell\in L_2(\ell)$ we have
\begin{align*}
\Thz_{\pb_i\pb_{i+1}}[w_\ell](\xb^\ell)&=
\frac{2\sqrt{3}}{3\ell}[w(\xb^\ell+\ell\pb_i-\ell \pb_{i+1})+\frac 32 \ell\gamma(\xb^\ell+\ell\pb_i-\ell \pb_{i+1})\\
&\hspace{1cm}-w(\xb^\ell+\ell \pb_{i})+w(\xb^\ell+\ell \pb_{i+1})-w(\xb^\ell)-\frac 32 \ell\gamma(\xb^\ell)],
\end{align*}
and by Taylor expanding $w$ up to second order and $\gamma$ up to first order, we find:
\begin{align*}
\Thz_{\pb_i\pb_{i+1}}[w_\ell](\xb^\ell)&=
\frac{2\sqrt{3}\ell}{3}\Big(\nabla^2w(\xb^\ell ) \pb_{i+1}\cdot ( \pb_{i+1}-\pb_i)\\
&\hspace{2cm}+\frac 32 \nabla \gamma(\xb^\ell)\cdot (\pb_i- \pb_{i+1})+o(1)\Big),\\
&={2\ell}\Big(\nabla^2w(\xb^\ell)\pb_{i+1}-\frac 32 \nabla \gamma(\xb^\ell)\Big)\cdot \frac{\pb_{i+1}-\pb_i}{|\pb_{i+1}-\pb_i|}+o(\ell).
\end{align*}
Similarly, we find:
$$
\Thz_{\pb_i\pb_{i+2}}[w_\ell](\xb^\ell)=\frac{2\sqrt{3}\ell}{3}\Big(\nabla^2w(\xb^\ell)\pb_{i+2}-\frac 32 \nabla \gamma(\xb^\ell)\Big)\cdot\frac{\pb_{i+2}-\pb_i}{|\pb_{i+2}-\pb_i|}+o(\ell).
$$
The $\mathcal{Z}$-dihedral energy \eqref{enZ} takes the form
\begin{align}\label{Uzl}
\mathcal{U}^{\mathcal{Z}}_\ell(w_\ell)=&2\ell^2k^{\mathcal{Z}}\, \sum_{\xb^\ell\in L_2(\ell)} \sum_{i=1}^3
\Big\{\Big((\nabla^2w(\xb^\ell)\pb_{i+1}-\frac 32 \nabla \gamma(\xb^\ell))\cdot \frac{\pb_{i+1}-\pb_i}{|\pb_{i+1}-\pb_i|}\Big)^2\nonumber \\
&+ \Big((\nabla^2w(\xb^\ell)\pb_{i+2}-\frac 32 \nabla \gamma(\xb^\ell))\cdot \frac{\pb_{i+2}-\pb_i}{|\pb_{i+2}-\pb_i|}\Big)^2\Big\}+o(1).
\end{align}
Let $E^\ell(\xb^\ell)$ be the hexagon of side length $\ell$ centered at $\xb^\ell$ and with two sides parallel to $\pb_1$,
see Fig. \ref{fig:hexagon}, and observe that the area of the hexagon $E^\ell(\xb^\ell)$ is $3\sqrt{3}\ell^2/2$. Thence, \eqref{Uzl} can be written as
$$
\mathcal{U}^{\mathcal{Z}}_\ell(w_\ell)=\frac{4\sqrt{3}}{9}k^{\mathcal{Z}}\int_{\Ro^2}W^{\mathcal{Z}}_\ell(\xb)\,d\xb+o(1),
$$
with $W^{\mathcal{Z}}_\ell$ defined by
\begin{align*}
W^{\mathcal{Z}}_\ell(\xb):=\sum_{\xb^\ell\in L_2(\ell)}& \sum_{i=1}^3
\Big\{\Big((\nabla^2w(\xb^\ell)\pb_{i+1}-\frac 32 \nabla \gamma(\xb^\ell))\cdot \frac{\pb_{i+1}-\pb_i}{|\pb_{i+1}-\pb_i|}\Big)^2\\
&+ \Big((\nabla^2w(\xb^\ell)\pb_{i+2}-\frac 32 \nabla \gamma(\xb^\ell))\cdot \frac{\pb_{i+2}-\pb_i}{|\pb_{i+2}-\pb_i|}\Big)^2\Big\}\chi_{E^\ell(\xb^\ell)}(\xb).
\end{align*}

Hence, by passing to the limit we find
$$
\lim_{\ell\to 0} \mathcal{U}_\ell(w_\ell)=\lim_{\ell\to 0}\mathcal{U}^{\mathcal{Z}}_\ell(w_\ell)=\frac{4\sqrt{3}}{9}k^{\mathcal{Z}}\int_{\Ro^2}W^{\mathcal{Z}}_0(\xb)\,d\xb,
$$
where
\begin{align*}
W^{\mathcal{Z}}_0(\xb):= \sum_{i=1}^3
\Big\{&\Big((\nabla^2w(\xb)\pb_{i+1}-\frac 32 \nabla \gamma(\xb))\cdot \frac{\pb_{i+1}-\pb_i}{|\pb_{i+1}-\pb_i|}\Big)^2\\
&+ \Big((\nabla^2w(\xb)\pb_{i+2}-\frac 32 \nabla \gamma(\xb))\cdot \frac{\pb_{i+2}-\pb_i}{|\pb_{i+2}-\pb_i|}\Big)^2\Big\}.
\end{align*}
From the definitions of $\pb_i$ and $\db_i$ it is 
\begin{equation}
\pb_2-\pb_1=\db_3, \quad \pb_3-\pb_2=\db_2, \quad \textrm{and} \quad \pb_1-\pb_3=\db_1.
\end{equation}
Then,  we easily check that the limit energy  coincides with $\mathcal{U}_0^{\mathcal{Z}}(w, \gamma)$.

We now prove $\it 2.$
Let 
$$
w_\ell(\xb):=\sum_{\xb^\ell\in L_1(\ell)\cup L_2(\ell)}w(\xb^\ell)\ \chi_{T^\ell(\xb^\ell)}(\xb).
$$
Then, $w_\ell \to w$ in $L^2(\Ro^2)$ and for $\ell$ small enough $w_\ell\in  \mathcal{A}_\ell$ . 
Setting $\gamma=0$ in the proof of $\it 1.$,
we find
\begin{equation}\label{Urecz}
\lim_{\ell\to 0}\mathcal{U}^{\mathcal{Z}}_\ell(w_\ell)=\mathcal{U}^{\mathcal{Z}}_0(w,0).
\end{equation}

Let us the consider the $\Cc$-dihedral energy. By Taylor expanding $w$ around $\xb^\ell$ up to second order, from \eqref{C-wedge} we find that
\begin{align*}
\Thc_{\pb_i^+}[w_\ell](\xb^\ell)&=2\ell\nabla^2w(\xb^\ell)\pb_{i}\cdot \pb_{i}^\perp+o(\ell),\\
\Thc_{\pb_i^-}[w_\ell](\xb^\ell)&=2\ell\nabla^2w(\xb^\ell)\pb_{i}\cdot \pb_{i}^\perp+o(\ell),
\end{align*}
see Appendix A.3 of \cite{Davini_2017} if further details are needed.
Then,   \eqref{enC} writes as
\begin{align*}
\mathcal{U}^{\mathcal{C}}_\ell(w_\ell)&=4 k^{\mathcal{C}} \, \sum_{\xb^\ell\in L_2(\ell)}
\sum_{i=1}^3 \Big(2\ell\nabla^2w(\xb^\ell)\pb_{i}\cdot \pb_{i}^\perp+o(\ell)\Big)^2\\
&=\frac{8\sqrt{3}}9 k^{\mathcal{C}} \int_{\Ro^2} \sum_{\xb^\ell\in L_2(\ell)}
\sum_{i=1}^3 \Big(\nabla^2w(\xb^\ell)\pb_{i}\cdot \pb_{i}^\perp\Big)^2\chi_{E^\ell(\xb^\ell)}(\xb)\,d\xb+o(1),
\end{align*}
where  $E^\ell(\xb^\ell)$ is the hexagon defined above of area $3\sqrt{3}\ell^2/2$.
From this identity we immediately deduce that
\begin{equation}\label{Urecc}
\lim_{\ell\to 0}\mathcal{U}^{\mathcal{C}}_\ell(w_\ell)=\mathcal{U}^{\mathcal{C}}_0(w).
\end{equation}

Similarly, from \eqref{s_angles} we find
\begin{align*}
\Ths_1[w_\ell](\xb^\ell)&=\frac{\sqrt{3\sqrt{3}}\ell}{6}\nabla^2w(\xb^\ell)\pb_{i}\cdot \pb_{i}+o(\ell),\\
\Ths_2[w_\ell](\xb^\ell)&=\frac{\sqrt{3\sqrt{3}}\ell}{6}\nabla^2w(\xb^\ell)\pb_{i}\cdot \pb_{i}+o(\ell),
\end{align*}
and hence the self-stress energy \eqref{ens} writes as
\begin{align*}
\mathcal{U}^{s}_\ell(w_\ell)&=-\frac{\sqrt{3}}{24} \tau_0\, \sum_{\xb^\ell\in L_1(\ell)\cup L_2(\ell)} 
\, \Big(\ell\,\nabla^2w(\xb^\ell)\pb_{i}\cdot \pb_{i}+o(\ell)\Big)^2\\
&=-\frac{1}{18} \tau_0\int_{\Ro^2} \sum_{\xb^\ell\in L_1(\ell)\cup L_2(\ell)} 
\, \Big(\nabla^2w(\xb^\ell)\pb_{i}\cdot \pb_{i}\Big)^2 \chi_{T^\ell(\xb^\ell)}(\xb)\,d\xb+o(1),
\end{align*}
since the area of $T^\ell(\xb^\ell)$ is $3\sqrt{3}\ell^2/4$.
It follows that
\begin{equation}\label{Urecs} 
\lim_{\ell\to 0}\mathcal{U}^{s}_\ell(w_\ell)=\mathcal{U}^{s}_0(w).
\end{equation}
From \eqref{Urecz}, \eqref{Urecc}, and \eqref{Urecs}, and recalling the definition \eqref{U00} of 
$\mathcal{U}_0$ we conclude the proof.
\QED

\noindent
{\sc Proof of Theorem  \ref{Gamma}.}
We first note that $\mathcal{U}_0$, as defined in \eqref{U00}, coincides with 
$\mathcal{U}^{(b)}_0$ as given in \eqref{U0b}. Indeed, it suffices to rewrite the derivatives appearing in $\mathcal{U}_0$ with respect to the coordinates $x_1$ and $x_2$ of a Cartesian orthogonal system,
see \cite{Davini_2017} for further details.
We need to prove that:
\begin{enumerate}
	\item (\textsc{Liminf inequality}) for every $w\in L^2(\Omega)$ and for every sequence $w_\ell$ converging to $w$ in $L^2(\Omega)$
	$$
	\liminf_{\ell\to 0} \mathcal{U}^{\rm e}_\ell (w_\ell)\ge \mathcal{U}^{\rm e}_0 (w);
	$$
	\item (\textsc{Recovery sequence}) for every $w\in L^2(\Omega)$ there exists a sequence $w_\ell$ converging to $w$ in $L^2(\Omega)$
	such that
	$$
	\limsup_{\ell\to 0} \mathcal{U}^{\rm e}_\ell (w_\ell)\le \mathcal{U}^{\rm e}_0 (w).
	$$
\end{enumerate}
We start by proving 1. Let $w, w_\ell\in L^2(\Omega)$ such that $w_\ell\to w$ in $L^2(\Omega)$ and,
without loss of generality, $\liminf_{\ell\to 0} \mathcal{U}^{\rm e}_\ell (w_\ell)<+\infty$.
Then, up to a subsequence (not relabeled), by \eqref{Uel} we have that 
$$
\mathcal{U}^{\rm e}_\ell (w_\ell)=\mathcal{U}_\ell (w_\ell), \quad \sup_\ell \mathcal{U}_\ell (w_\ell)<+\infty, \quad w_\ell\in \mathcal{A}_\ell.
$$
By Lemma \ref{regularity}, $w\in H^2_0(\Omega)$, and by Theorem \ref{gamma} we find $\gamma=0$. By combining Lemmas \ref{thm:liminfZ}, \ref{thm:liminfC}, and \ref{thm:liminfs}, we deduce that
$$
\liminf_{\ell\to 0} \mathcal{U}_\ell (w_\ell)\ge \mathcal{U}_0 (w).
$$

We now prove 2. Let $w\in L^2(\Omega)$ be such that,
without loss of generality, $\mathcal{U}^{\rm e}_0 (w)<+\infty$.
Then, from the definition of $\mathcal{U}^{\rm e}_0$ we infer that $w\in H^2_0(\Omega)$.
Let $w^k\in C_0^\infty(\Omega)$ be a sequence such that $w^k\to w$ in $H^2(\Omega)$
as $k$ tends to $+\infty$, so that
$$
\lim_{k\to+\infty} \mathcal{U}_0 (w^k)= \mathcal{U}_0 (w).
$$
By Lemma \ref{rec0}, for every $k$ there exists a sequence $w^k_\ell$ such that $w^k_\ell\to w^k$ in $L^2(\Omega)$, as $\ell\to 0$, and
$$
\limsup_{\ell\to 0} \mathcal{U}_\ell (w^k_\ell)\le \mathcal{U}_0 (w^k).
$$
Combining the two limits we find
$$
\lim_{k\to+\infty}\limsup_{\ell\to 0} \mathcal{U}_\ell (w^k_\ell)\le\lim_{k\to+\infty} \mathcal{U}_0 (w^k)= \mathcal{U}_0 (w).
$$
By a diagonal argument there exists an increasing mapping $\ell\mapsto k(\ell)$ such that
$w^{k(\ell)}_\ell\to w$ in $L^2(\Omega)$ and 
$$
\limsup_{\ell\to 0} \mathcal{U}_\ell (w^{k(\ell)}_\ell)\le \mathcal{U}_0 (w).
$$
Hence, part 2. is proven.
\QED

{\sc Proof of Theorem  \ref{Gamma2}.}
We recall that $\mathcal{U}^{\mathcal{Z}}_0$ defined in \eqref{Uz00} takes also the form given in \eqref{Uz0}. 

We start by proving the liminf inequality. Let $w, w_\ell\in L^2(\Omega)$ such that $w_\ell\to w$ in $L^2(\Omega)$. Arguing as in the proof of Theorem \ref{Gamma},  from the assumption that $\sup_\ell \mathcal{U}^{\mathcal{Z}\rm e}_\ell<\infty$ we deduce that $
\mathcal{U}^{\mathcal{Z}\rm e}_\ell (w_\ell)=\mathcal{U}^{\mathcal{Z}}_\ell (w_\ell)$, that $w_\ell\in \mathcal{A}_\ell$
and that $w\in H^2_0(\Omega)$. By Lemma \ref{thm:liminfZ},  it follows that
\begin{align}\label{liminf}
\liminf_{\ell\to 0} \mathcal{U}^{\mathcal{Z}}_\ell (w_\ell)\ge \mathcal{U}^{\mathcal{Z}}_0 (w,\gamma)&\ge \inf_{\gamma\in H^1_0(\Omega)}\mathcal{U}^{\mathcal{Z}}_0 (w,\gamma)\nonumber\\
&=\mathcal{U}^{\mathcal{Z}}_0\left(w,(-\Delta)^{-1}(-\tfrac 23 \partial_{\pb_1\pb_2\pb_3}w)\right),
\end{align}
where the last identity can be found by writing the Euler-Lagrange equation that  the minimizer $\gamma$ satisfies.

We now prove the recovery sequence condition. Without loss of generality, let $w\in L^2(\Omega)$ be such that $\mathcal{U}^{\mathcal{Z}\rm e}_0 (w)<+\infty$.
Then, from the definition of $\mathcal{U}^{\mathcal{Z}\rm e}_0$ we infer that $w\in H^2_0(\Omega)$.
Set
$$
\gamma:=(-\Delta)^{-1}(-\tfrac 23 \partial_{\pb_1\pb_2\pb_3}w)\in H^1_0(\Omega).
$$
Let $w^k, \gamma^k\in C_0^\infty(\Omega)$ be two sequences such that $w^k\to w$ in $H^2(\Omega)$
and $\gamma^k\to \gamma$ in $H^1(\Omega)$,
as $k$ tends to $+\infty$. Then, 
$$
\lim_{k\to+\infty} \mathcal{U}^{\mathcal{Z}}_0 (w^k,\gamma^k)= \mathcal{U}^{\mathcal{Z}}_0 (w,\gamma).
$$
By Lemma \ref{rec0}, for every $k$ there exists a sequence $w^k_\ell$ such that $w^k_\ell\to w^k$ in $L^2(\Omega)$, as $\ell\to 0$, and
\begin{equation}\label{b}
\limsup_{\ell\to 0} \mathcal{U}^{\mathcal{Z}}_\ell (w^k_\ell)\le \mathcal{U}^{\mathcal{Z}}_0 (w^k,\gamma^k).
\end{equation}
So, passing to the limit on the two sides of \eqref{b} yields that
$$
\lim_{k\to+\infty}\limsup_{\ell\to 0} \mathcal{U}^{\mathcal{Z}}_\ell (w^k_\ell)\le \lim_{k\to+\infty}\mathcal{U}^{\mathcal{Z}}_0 (w^k,\gamma^k)=\mathcal{U}^{\mathcal{Z}}_0 (w,\gamma).
$$
Again, by a diagonal argument there exist $w^{k(\ell)}_\ell$ such that
$w^{k(\ell)}_\ell\to w$ in $L^2(\Omega)$ and 
\begin{equation}\label{limsup}
\limsup_{\ell\to 0} \mathcal{U}_\ell (w^{k(\ell)}_\ell)\le \mathcal{U}^{\mathcal{Z}}_0 (w,(-\Delta)^{-1}(-\tfrac 23 \partial_{\pb_1\pb_2\pb_3}w)),
\end{equation}
which completes the proof.
\QED

\section*{Acknowledgments}	
AF acknowledges the financial support of Sapienza University of Rome (Progetto d'Ateneo 2016 --- ``Multiscale Mechanics of 2D Materials: Modeling and Applications''). 

\bibliographystyle{plain}
\bibliography{bibtex}

\addcontentsline{toc}{section}{References}

\end{document}